\documentclass[amssymb,prb,aps,twocolumn,amsmath,superscriptaddress,showpacs]{revtex4}

\usepackage{graphicx}

\begin{document}

\title{Dephasing of Electrons in Mesoscopic Metal Wires}
\author{F. Pierre}
\email{fred.pierre@laposte.net} \altaffiliation[Permanent address after January 01, 2004: ]{Laboratoire de Photonique
et de Nanostructures (LPN)-CNRS, route de Nozay, 91460 Marcoussis, France.} \affiliation{Department of Physics and
Astronomy, Michigan State University, East Lansing, Michigan 48824-2320, USA} \affiliation{Service de Physique de
l'Etat Condens\'{e}, Direction des Sciences de la Mati\`{e}re, CEA-Saclay, 91191 Gif-sur-Yvette, France}
\affiliation{Department of Applied Physics, Yale University, New Haven, Connecticut 06520, USA}
\author{A.B. Gougam}
\altaffiliation[Present address: ]{Centre for Advanced Nanotechnolgy, University of Toronto, Toronto, Ontario M5S 3E3,
Canada.} \affiliation{Department of Physics and Astronomy, Michigan State University, East Lansing, Michigan
48824-2320, USA}
\author{A. Anthore}
\author{H. Pothier}
\author{D. Esteve}
\affiliation{Service de Physique de l'Etat Condens\'{e}, Direction des Sciences de la Mati\`{e}re, CEA-Saclay, 91191
Gif-sur-Yvette, France}
\author{Norman O. Birge}
\affiliation{Department of Physics and Astronomy, Michigan State University, East Lansing, Michigan 48824-2320,
USA}
\date{\today}

\begin{abstract}
We have extracted the phase coherence time $\tau_{\phi}$ of electronic quasiparticles from the low field
magnetoresistance of weakly disordered wires made of silver, copper and gold. In samples fabricated using our purest
silver and gold sources, $\tau_{\phi}$ increases as $T^{-2/3}$ when the temperature $T$ is reduced, as predicted by the
theory of electron-electron interactions in diffusive wires. In contrast, samples made of a silver source material of
lesser purity or of copper exhibit an apparent saturation of $\tau_{\phi}$ starting between 0.1 and 1~K down to our
base temperature of 40~mK. By implanting manganese impurities in silver wires, we show that even a minute concentration
of magnetic impurities having a small Kondo temperature can lead to a quasi saturation of $\tau_{\phi}$ over a broad
temperature range, while the resistance increase expected from the Kondo effect remains hidden by a large background.
We also measured the conductance of Aharonov-Bohm rings fabricated using a very pure copper source and found that the
amplitude of the $h/e$ conductance oscillations increases strongly with magnetic field. This set of experiments
suggests that the frequently observed ``saturation'' of $\tau_{\phi}$ in weakly disordered metallic thin films can be
attributed to spin-flip scattering from extremely dilute magnetic impurities, at a level undetectable by other means.
\end{abstract}

\pacs{73.23.-b, 73.50.-h, 71.10.Ay, 72.70.+m} \maketitle
\section{Motivations}

The time $\tau_{\phi}$ during which the quantum coherence of an electron is maintained is of fundamental importance in
mesoscopic physics. The observability of many phenomena specific to this field relies on a long enough phase coherence
time.\cite{meso} Amongst these are the weak localization correction to the conductance (WL), the universal conductance
fluctuations (UCF), the Aharonov-Bohm (AB) effect, persistent currents in rings, the proximity effect near the
interface between a superconductor and a normal metal, and others. Hence it is crucial to find out what mechanisms
limit the quantum coherence of electrons.

In metallic thin films, at low temperature, electrons experience mostly elastic collisions from sample boundaries,
defects of the ion lattice and static impurities in the metal. These collisions do not destroy the quantum coherence of
electrons. Instead they can be pictured as resulting from a static potential on which the diffusive-like electronic
quantum states are built.

What limits the quantum coherence of electrons are inelastic collisions. These are collisions with other electrons
through the screened Coulomb interaction, with phonons, and also with extrinsic sources such as magnetic impurities or
two level systems in the metal. Whereas above about 1~K electron-phonon interactions are known to be the dominant
source of decoherence,\cite{phonons} electron-electron interactions are expected to be the leading inelastic process at
lower temperatures in samples without extrinsic sources of decoherence.\cite{AAK}

The theory of electron-electron interactions in the diffusive regime was worked out in the early 1980's (for a review
see \cite{AA}). It predicts a power law divergence of $\tau_{\phi}$ when the temperature $T$ goes to zero. Effects of
quantum interference are therefore expected to grow significantly upon cooling down the electrons. In mesoscopic wires,
the predicted power law $\tau_{\phi}\propto T^{-2/3}$ was first observed in 1986 by Wind \emph{et al.}\cite{Wind}
between 2~K and 5~K in aluminum and silver wires and then by Echternach \emph{et al.}\cite{Echternach} down to 100~mK
in a gold wire. However, in 1997, Mohanty, Jariwala, and Webb\cite{MJW} published a series of measurements of
$\tau_{\phi}$ on gold wires with a broad range of diffusion coefficients. They observed that the phase coherence time
tends to saturate at low temperature, typically below 0.5 K, in apparent contradiction with theoretical predictions.
That same year, measurements of the energy exchange rate between electrons in copper wires\cite{PRLrelax} were found to
be at odds, both qualitatively and quantitatively, with the prediction for electron-electron interactions. Both
experiments suggested that electrons in mesoscopic metallic wires interact with each other differently and more
strongly than predicted by theory.

To shed some light on this issue we present here several sets of experiments probing the phase coherence time at low
temperature in mesoscopic metal wires.\cite{names} We summarize our most important conclusions here. First, we measured
$\tau_{\phi}(T)$ down to 40~mK in several wires made of copper, silver, and gold and fabricated from source materials
of various purities. We found in the four very pure silver wires and in the very pure gold wire that $\tau_{\phi}(T)$
does not saturate in the investigated temperature range, but continues to increase as the temperature is lowered in
agreement with the theoretical prediction. Since these samples have comparable resistances and geometries as some
measured in \cite{MJW}, this observation casts doubt on the assertion \cite{MJW} that saturation of $\tau_{\phi}$ is a
universal feature of weakly-disordered metals. Second, we tested the impact of very dilute magnetic impurities with a
small Kondo temperature on the temperature dependence of $\tau_{\phi}$. We found that even at concentrations lower than
one part per million (1~ppm), such impurities can cause $\tau_{\phi}(T)$ to display a plateau over a large temperature
range. This could explain why saturation of $\tau_{\phi}$ at low temperature is frequently observed. Finally, we probed
the magnetic field dependence of the phase coherence time by measuring the magnetoresistance of copper Aharonov-Bohm
rings showing a temperature-independent $\tau_{\phi}$ at low temperature. The amplitude of the Aharonov-Bohm
conductance oscillations increased strongly on a field scale proportional to the temperature, indicating that the phase
coherence time at zero field was limited by spin-flip scattering from magnetic impurities.

\section{Experimental techniques}

\subsection{Sample fabrication}

\begin{figure}[ptbh]
\includegraphics[width=2.3in]{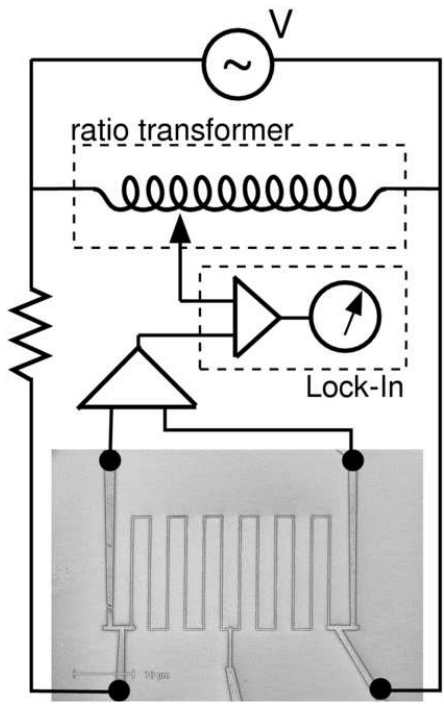} \caption{Photograph of a
silver sample taken with a scanning electron microscope, and schematic of measurement circuit. The wire resistance is
obtained by a four-lead measurement in a bridge configuration: the current is injected by two arms through the bias
resistor and the voltage is measured across two other arms in order to probe only the wire resistance; a ratio
transformer is used to enhance sensitivity to small variations of the sample resistance.} \label{FigSampleWL}
\end{figure}

Figure~\ref{FigSampleWL} displays the photograph of a typical sample together
with a schematic of the measurement setup.

All samples were fabricated using standard e-beam lithography techniques. A bilayer resist, consisting of a copolymer
P(MMA/MAA) bottom layer and a PMMA top layer, was first spun onto an oxidized Si substrate wafer. This bilayer was then
patterned by e-beam lithography to tailor a mask. The metal --- gold, copper or silver --- was deposited directly
through this mask in evaporators used only for non-magnetic metals.\cite{adhesion}

Samples made at Saclay used a Si substrate thermally oxidized over 500~nm, and
metal evaporation was performed in an electron gun evaporator. The silver
source material was placed inside a carbon liner, whereas copper and gold were
put directly in the buckets of the e-gun system. Metal evaporation took place
at a base pressure of about $10^{-6}$~mbar with an evaporation rate of 0.2,
0.5 and 1~nm/s for silver, gold and copper respectively (see\cite{PierrePHD}).

Samples made at Michigan State University (MSU) were evaporated on a Si substrate with only the native oxide in a
thermal evaporator used only for silver, aluminum, gold, copper and titanium. The source material and boat were
replaced before each evaporation and manipulated using plastic tweezers. The pressure during evaporation was about
$10^{-6}$~mbar and the evaporation rate ranged between 0.2 and 0.5~nm/s.\cite{NoteEVAP}

We measured the low field magnetoresistance of copper, gold and silver wires fabricated using source materials of
purity 99.999\% (5N) and 99.9999\% (6N). Electrical and geometrical characteristics of the samples are summarized in
Table~\ref{tableParam}.

\begin{table}[ptbh]
\begin{tabular}
[c]{|c|c|c|c|c|c|c|}\hline
Sample & Made & $L$ & $t$ & $w$ & $R$ & $D$\\
& at & ($\mu$m) & (nm) & (nm) & (k$\Omega$) & ($\mathrm{cm}^{2}/\mathrm{s}%
$)\\\hline
Ag(6N)a & Saclay & 135 & 45 & 65 & 1.44 & 115\\
Ag(6N)b & Saclay & 270 & 45 & 100 & 3.30 & 70\\
Ag(6N)c & Saclay & 400 & 55 & 105 & 1.44 & 185\\
Ag(6N)d & MSU & 285 & 35 & 90 & 1.99 & 165\\
Ag(5N)a & Saclay & 135 & 65 & 108 & 0.68 & 105\\
Ag(5N)b & Saclay & 270 & 65 & 90 & 1.31 & 135\\
Ag(5N)c$_{\mathrm{Mn}0.3}$ & Saclay & 135 & 65 & 110 & 0.47 & 150\\
Ag(5N)d$_{\mathrm{Mn}1}$ & Saclay & 270 & 65 & 95 & 1.22 & 135\\
Au(6N) & MSU & 175 & 45 & 90 & 1.08 & 135\\
Cu(6N)a & MSU & 285 & 45 & 155 & 0.70 & 145\\
Cu(6N)b & MSU & 285 & 20 & 70 & 7.98 & 60\\
Cu(6N)c & MSU & 285 & 35 & 75 & 4.37 & 65\\
Cu(6N)d & MSU & 285 & 20 & 80 & 8.50 & 50\\
Cu(5N)a & Saclay & 270 & 45 & 110 & 1.68 & 70\\
Cu(5N)b & Saclay & 270 & 45 & 100 & 0.95 & 160\\\hline
\end{tabular}
\newline \caption{Geometrical and electrical characteristics of the measured
samples.\cite{correspondance} The diffusion coefficient $D$ is obtained using Einstein's relation
$1/\rho=\nu_{F}e^{2}D$ with the density of states in copper, silver and gold respectively $\nu_{F}=1.56\times10^{47}$,
$1.03\times10^{47}$ and $1.14\times10^{47}~\mathrm{J^{-1}m^{-3}}$, and the resistivity $\rho$ extracted from the
resistance $R$, thickness $t$, length $L$ and width $w$ of the long wire. Length and width were measured with a
scanning electron microscope (SEM). The thickness of most samples was measured with an atomic force microscope (AFM);
for others the value given by a calibrated thickness monitor in the evaporator was used. A rectangular cross-section is
assumed.} \label{tableParam}
\end{table}

\subsection{Experimental setup}

The samples were immersed in the mixing chamber of a top loading dilution refrigerator. Electrical lines to the sample
were filtered by commercial ``pi'' filters at the top of the cryostat and by discrete RC filters in the mixing chamber.
Resistance measurements were performed using a standard ac four-terminal technique with a room temperature
pre-amplifier of input voltage noise 1.5~nV/$\sqrt{\mathrm{Hz}}$ and a lock-in amplifier operated at frequencies
between 100 and 300~Hz (see Fig.~\ref{FigSampleWL}). To avoid significant heating of electrons we used ac voltages
$V_{ac}$ across the samples such that $eV_{ac}\lesssim k_{B}T$. This is particularly important at temperatures below
100~mK for which the length scale for electron-phonon interactions, responsible for cooling down the electronic fluid,
can be as large as several millimeters (see Appendix A). A bridge circuit with a ratio transformer on one arm was used
to enhance the measurement sensitivity to small changes in sample resistance. A magnetic field was applied
perpendicular to the plane of the sample using a superconducting coil.

\section{Low field magnetoresistance measurements}

The most accurate way to extract $\tau_{\phi}$ at low magnetic field in metallic thin films is to measure the
magnetoresistance and to fit it using weak localization theory.\cite{WLreview} Both the amplitude and width of the weak
localization peak (dip when spin-orbit coupling is strong) in the resistance are sensitive to the phase coherence
length.

\begin{figure}[ptbh]
\includegraphics[width=3.4in]{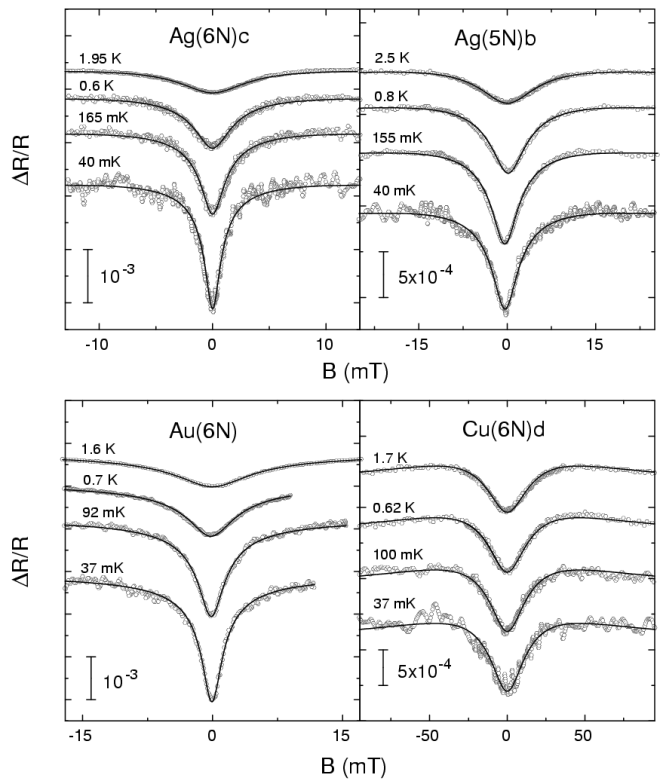} \caption{Magnetoresistance data
(symbols) and fits to equation~(\ref{WeakLoc2}) (solid lines). Top panels are measurements of two silver samples made
of source materials of nominal purity 6N (99.9999\%, top left panel) and 5N (99.999\%, top right panel). Bottom panels
display data measured on gold (bottom left panel) and copper (bottom right panel) samples made of 6N nominal purity
source materials. The curves are offset vertically for clarity.} \label{FigRawWL}
\end{figure}

Figure~\ref{FigRawWL} displays the low field magnetoresistance of samples
Ag(6N)c, Ag(5N)b, Au(6N) and Cu(6N)d at several temperatures. The positive
magnetoresistance indicates that spin-orbit scattering is stronger than
inelastic scattering ($\tau_{so}<\tau_{\phi}$). Raw magnetoresistance
measurements already reveal a qualitative difference between these samples:
the dip in the magnetoresistance of samples Ag(6N)c and Au(6N) becomes deeper
and narrower upon cooling down to base temperature whereas it stops changing
at low temperature in samples Ag(5N)b and Cu(6N)d.

The magnetoresistance $\Delta R \equiv R(B)-R(\infty)$ is fit with the quasi-1D expression for the weak
localization correction
\begin{eqnarray}
\frac{\Delta R}{R}  &=& \frac{2R}{R_{\mathrm{K}}L}\left\{  \frac{3}{2}\left[\frac{1}{L_{\phi}^{2}}+\frac{4}{3L_{so}^{2}}+\frac{1}{3}\left(  \frac{w}{L_{H}^{2}}\right)  ^{2}\right]  ^{-1/2}\right. \nonumber\\
&&   \left.  -\frac{1}{2}\left[  \frac{1}{L_{\phi}^{2}}+\frac{1}{3}\left( \frac{w}{L_{H}^{2}}\right)  ^{2}\right]
^{-1/2}\right\} \label{WeakLoc2}
\end{eqnarray}
where $R$ is the resistance of a wire of length $L$ and width $w$, $R_{\mathrm{K}}=h/e^{2}$ is the resistance quantum,
$L_{\phi}=\sqrt {D\tau_{\phi}}$ is the phase coherence length, $D$ is the diffusion coefficient of electrons,
$L_{H}=\sqrt{\hbar/eB}$ is the magnetic length, $B$ is the magnetic field applied perpendicularly to the sample plane,
and $L_{so}=\sqrt{D\tau_{so}}$ is the spin-orbit length that characterizes the intensity of spin-orbit coupling.
Expression~(\ref{WeakLoc2}) holds for metallic wires in the diffusive regime, far from the metal-insulator
transition, and in the quasi-1D regime: $l_{e}\ll w,t\ll L_{H},L_{\phi}%
,L_{so}\ll L$, with $t$ the sample thickness and $l_{e}$ the elastic mean free
path of electrons (see\cite{AAwl,AleinerWav} and Appendix~B).

In the fit procedure, we use the measured sample resistance and length given in Table~\ref{tableParam}. Our
experimental setup being designed to measure resistance changes with an higher accuracy than absolute values, $\Delta
R$ is known only up to a small additive constant that we adjusted to fit each magnetoresistance curve. The width was
fixed at a value $w_{\mathrm{WL}}$ giving the best overall fits for the complete set of data at various temperatures.
The difference between the width $w$ measured from scanning electron microscope images and the best fit value
$w_{\mathrm{WL}}$ (see Table~\ref{tableWLfit}) was found to be always less than 15\%.\cite{NoteWidth} The spin-orbit
length $L_{so}$ was obtained from fits of the magnetoresistance measured at the highest temperature. These parameters
being determined, $L_{\phi}$ remains as the only fit parameter for each magnetoresistance curve. Examples of fits are
displayed as solid lines in Fig.~\ref{FigRawWL}.

\begin{figure}[ptbh]
\includegraphics[width=3.1in]{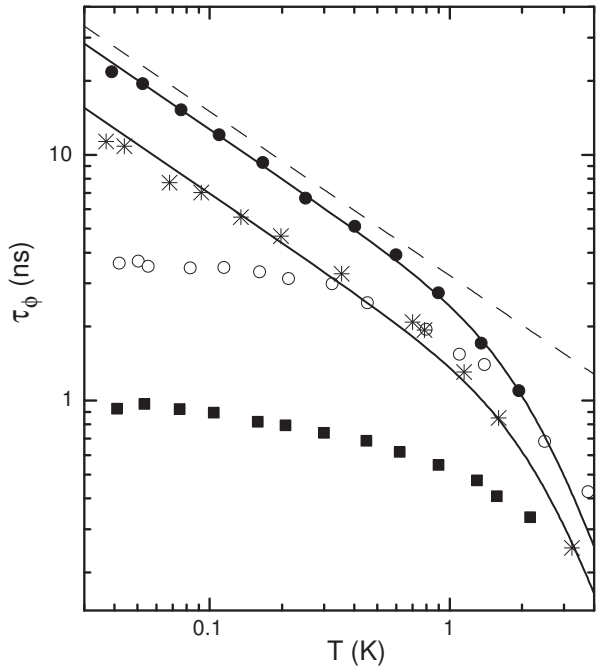} \caption{Phase coherence time
$\tau_{\phi}$ versus temperature in wires made of copper Cu(6N)b ($\blacksquare$), gold Au(6N) ($\ast$), and silver
Ag(6N)c ($\bullet$) and Ag(5N)b ($\circ$). The phase coherence time increases continuously with decreasing temperature
in wires fabricated using our purest (6N) silver and gold sources as illustrated respectively with samples Ag(6N)c and
Au(6N). Continuous lines are fits of the measured phase coherence time including inelastic collisions with electrons
and phonons (Eq.~(\ref{fitee-eph})). The dashed line is the prediction of electron-electron interactions only
(Eq.~(\ref{tauEEtext})) for sample Ag(6N)c. In contrast, the phase coherence time increases much more slowly in samples
made of copper (independently of the source material purity) and in samples made of silver using our source of lower
(5N) nominal purity.} \label{FigTauFi}
\end{figure}

In order to get $\tau_{\phi}$ from $L_{\phi},$ the diffusion coefficient $D$ was determined using the measured
geometrical and electrical sample characteristics given in Table~\ref{tableParam}. Figure~\ref{FigTauFi} shows
$\tau_{\phi}$ as a function of temperature for samples Ag(6N)c, Ag(5N)b, Au(6N) and Cu(6N)b. This confirms
quantitatively the differences between samples already mentioned from the raw magnetoresistance data. On the one hand,
the samples Ag(6N)c and Au(6N), fabricated using our purest (6N) silver and gold sources, present a large phase
coherence time that keeps increasing at low temperature. On the other hand, the copper sample Cu(6N)b and the sample
Ag(5N)b, fabricated using a silver source of smaller nominal purity (5N), present a much smaller phase coherence time
and exhibit a plateau in $\tau_{\phi}(T)$, in contradiction with the theoretical prediction for electron-electron
interactions. This trend, illustrated in Fig.~\ref{FigTauFi}, has been confirmed by the measurements of several
samples, as summarized in Table~\ref{tableWLfit}.

\begin{table}[ptbh]
\begin{tabular}
[c]{|c|c|c|c|c|}\hline
Sample & $\tau_{\phi}^{\mathrm{max}}$ & $L_{so}$ & $w_{\mathrm{WL}}$ ($w$)\\
& (ns) & ($\mu$m) & (nm) \\\hline
Ag(6N)a & $9\nearrow$ & 0.65 & 57 (65)\\
Ag(6N)b & $12\nearrow$ & 0.35 & 85 (100)\\
Ag(6N)c & $22\nearrow$ & 1.0 & 90 (105)\\
Ag(6N)d & $12\nearrow$ & 0.82 & 75 (90)\\
Ag(5N)a & 2.9 & 0.65 & 108 (108)\\
Ag(5N)b & 3.5 & 0.75 & 82 (90)\\
Au(6N) & $11\nearrow$ & 0.085 & 85 (90)\\
Cu(6N)a & 0.45 & 0.67 & 155 (155)\\
Cu(6N)b & 0.95 & 0.4 & 70 (70)\\
Cu(6N)c & 0.2 & 0.35 & 75 (75)\\
Cu(6N)d & 0.35 & 0.33 & 80 (80)\\
Cu(5N)a & 1.8 & 0.52 & 110 (110)\\
Cu(5N)b & 0.9 & 0.67 & 100 (100)\\\hline
\end{tabular}
\caption{Fit parameters of the magnetoresistance data to weak localization theory: maximum phase coherence time
$\tau_{\phi}^{\mathrm{max}},$ obtained at the lowest temperature of $\sim 40$~mK; spin orbit length $L_{so}$ and
effective width $w_{\mathrm{WL}}$. We also recall the width $w$ obtained from SEM pictures. The upwards arrow
$\nearrow$ indicates that $\tau_{\phi}$ keeps increasing down to 40~mK. In the other samples, $\tau_{\phi}$ is nearly
constant at low temperature.} \label{tableWLfit}
\end{table}

\section{Comparison with theoretical predictions - discussion}

\subsection{Purest silver and gold samples}

Theory predicts that, in samples without extrinsic sources of decoherence, $\tau_{\phi}(T)$ is limited by the
contributions of electron-electron $\tau_{\mathrm{ee}}$ and electron-phonon $\tau_{\mathrm{ph}}$ interactions. In
principle, decoherence by electron-electron scattering is not purely an exponential process, hence the decoherence
rates from electron-electron and electron-phonon scattering do not simply add. In practice (see Appendix B), the exact
formula for the magnetoresistance is indistinguishable from Eq.~(\ref{WeakLoc2}) with a total decoherence rate:
\begin{equation}
\frac{1}{\tau_{\phi}(T)}=\frac{1}{\tau_{\mathrm{ee}}(T)}+\frac{1}{\tau_{\mathrm{ph}}(T)}.\label{tauee}
\end{equation}

\begin{figure}[ptbh]
\includegraphics[width=3.2in]{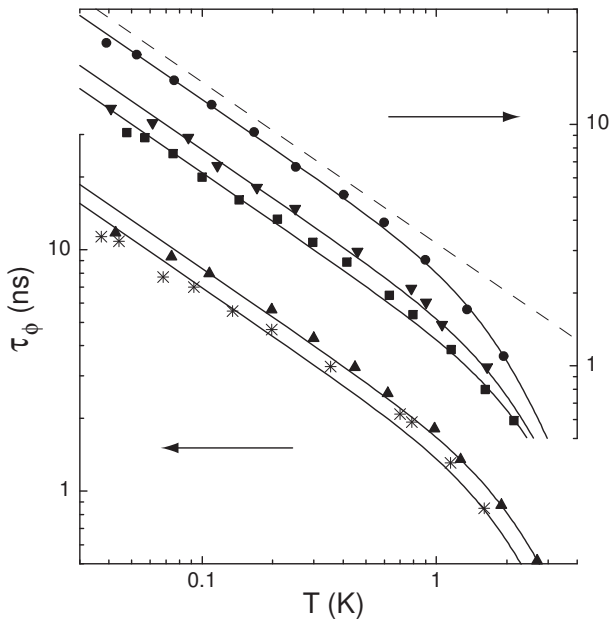} \caption{Phase coherence time vs
temperature in samples Ag(6N)a ($\blacksquare$), Ag(6N)b ($\blacktriangledown$), Ag(6N)c ($\bullet$), Ag(6N)d
($\blacktriangle$), and Au(6N) ($\ast$), all made of 6N sources. Continuous lines are fits of the data to
Eq.~(\ref{fitee-eph}). For clarity, the graph has been split in two part, shifted vertically one with respect to the
other. The quantitative prediction of Eq.~(\ref{tauEEtext}) for electron-electron interactions in sample Ag(6N)c is
shown as a dashed line.} \label{FigAg6N}
\end{figure}

For our wires, whose width and thickness are smaller than $L_{\phi}$, the
quasi-1D prediction for electron-electron interactions
applies\cite{AleinerWav}
\begin{equation}
\tau_{\mathrm{ee}}=\hbar\left[  \frac{(4/\pi)(R_{\mathrm{K}}/R)\nu_{F} SL}{(k_{B}T)^{2}}\right]
^{1/3}\equiv\frac{1}{A_{thy}T^{2/3}} ,\label{tauEEtext}
\end{equation}
where $\nu_{F}$ is the density of states per unit volume at the Fermi energy, and $S$ is the cross section of the wire.

In order to test the theoretical predictions, the measured $\tau_{\phi}(T)$ curves were fit with the functional form
\begin{equation}
\tau_{\phi}^{-1}=AT^{2/3}+BT^{3},\label{fitee-eph}
\end{equation}
where the second term describes electron-phonon scattering, dominant at higher temperatures.\cite{phonons} Fits are
shown as continuous lines in Fig.~\ref{FigAg6N} (the fit parameters minimize the distance between the data points and
the fit curve in a log-log plot). Equation~(\ref{fitee-eph}) describes accurately the temperature dependence of
$\tau_{\phi}(T)$ for samples Ag(6N)a, b, c and, with a slightly reduced fidelity, for samples Ag(6N)d and sample
Au(6N). In all these samples, fabricated using 6N source materials of silver and gold, $\tau_{\phi}(T)$ follows very
closely, below about 1~K, the $1/T^{2/3}$ dependence expected when electron-electron interaction is the dominant
inelastic process. Nevertheless, if the exponent of $T$ is left as a fit parameter, better fits are obtained with
smaller exponents ranging from $0.59$ for samples Ag(6N)d and Au(6N) up to $0.64$ for sample Ag(6N)c. This issue will
be discussed in part V.B. The dashed line in Fig.~\ref{FigTauFi} and Fig.~\ref{FigAg6N} is the quantitative prediction
of Eq.~(\ref{tauEEtext}) for electron-electron interactions in sample Ag(6N)c. The dephasing times are close, though
always slightly smaller, to the theoretical prediction of Eq.~(\ref{tauEEtext}). Table~\ref{tableFitEE} lists the best
fit parameters $A$, $B$, together with the prediction $A_{thy}$ of Eq.~(\ref{tauEEtext}).

\begin{table}[ptbh]
\begin{tabular}
[c]{|c|c|c|c|}\hline
Sample & $A_{thy}$ & $A$ & $B$\\
& (ns$^{-1}$K$^{-2/3}$) & (ns$^{-1}$K$^{-2/3}$) & (ns$^{-1}$K$^{-3}$)\\\hline
Ag(6N)a & 0.55 & 0.73 & 0.045\\
Ag(6N)b & 0.51 & 0.59 & 0.05\\
Ag(6N)c & 0.31 & 0.37 & 0.047\\
Ag(6N)d & 0.47 & 0.56 & 0.044\\
Au(6N) & 0.40 & 0.67 & 0.069\\\hline
\end{tabular}
\caption{Theoretical predictions of Eq.~(\ref{tauEEtext}) and fit parameters for $\tau_{\phi}(T)$ in the purest silver
and gold samples using the functional form given by Eq.~(\ref{fitee-eph}).} \label{tableFitEE}
\end{table}

This data set casts doubt on the claim by Mohanty, Jariwala and Webb\cite{MJW} (MJW) that saturation of $\tau_{\phi}$
is a universal phenomenon in mesoscopic wires. One can always argue that the saturation temperature for our silver
samples is below 40~mK, hence unobservable in our experiments. However, the resistivity and dimensions of sample
Ag(6N)a are similar to those of sample Au-3 in the MJW paper,\cite{MJW} which exhibits saturation of $\tau_{\phi}$
starting at about 100~mK, and has a maximum value of $\tau_{\phi}^{\mathrm{max}}=2$~ns. In contrast, $\tau_{\phi}$
reaches 9~ns in Ag(6N)a.

\subsection{Silver 5N and copper samples}

In silver samples made from a 5N purity source, the phase coherence time is systematically shorter than predicted by
Eq.~(\ref{tauEEtext}) and displays an unexpectedly flat temperature dependence below 400~mK. The same is true for all
the copper samples we measured, independently of source purity.\cite{NoteAu} These trends are illustrated for samples
Ag(5N)b and Cu(6N)b in Fig.~\ref{FigTauFi}.

What can be responsible for this anomalous behavior? There have been several theoretical suggestions regarding sources
of extra dephasing. Some of these, such as the presence of a parasitic high frequency electromagnetic
radiation,\cite{AAG} can be ruled out purely on experimental grounds. Indeed some samples do show a saturation of
$\tau_{\phi}$, while others of similar resistance and geometry, measured in the same cryostat, do not. This indicates
that, in our experiments at least, the observed excess dephasing is not an artifact of the measurement. The main
suggestions to explain the anomalous behavior of $\tau_{\phi}$ are dephasing by very dilute magnetic
impurities,\cite{dephasingMI,PierrePHD} dephasing by two-level systems associated with lattice defects, \cite{IFS,2CK}
and dephasing by electron-electron interactions through high energy electromagnetic modes.\cite{GZ}

The correlation between source material purity and excess dephasing amongst silver samples fabricated using the exact
same process but with either our 5N or 6N source material suggests that impurities are responsible for the anomalous
temperature dependence of $\tau_{\phi}$. The fact that, among all the 6N silver samples, $\tau_{\phi}(T)$ deviates the
most from the prediction of electron-electron interactions in Ag(6N)d, fabricated in MSU (see Fig.~\ref{FigAg6N}) would
mean that the 6N silver source material used at MSU contains more ``dangerous'' impurities than the one at Saclay.

The phase coherence time in the copper samples is always almost independent of temperature below about 200~mK down to
our base temperature of 40~mK (see \cite{Gougam,PierrePHD,PierreAB}). However, as opposed to silver samples, this
unexpected behavior is not correlated with the source material purity (5N or 6N). A likely explanation is provided by
early measurements showing that the surface oxide of copper can cause dephasing.\cite{Haesendonck}

\section{Influence on $\tau_\phi$ of very dilute magnetic impurities}

Dephasing of conduction electrons by paramagnetic impurities has been known since 1980,\cite{dephasingMI}
hence it may come as a surprise that this issue is still under debate today. In their Letter on the ``saturation'' of
$\tau_{\phi}$ at low temperature,\cite{MJW} Mohanty, Jariwala, and Webb studied the effect of intentionally doping
their gold wires with iron impurities. They found that $\tau_{\phi}$ in those samples did not truly saturate, but
rather reached a plateau around 1~K and increased again below about 0.3~K. In addition, the presence of the iron
impurities could be detected by a logarithmic contribution to the temperature dependence of the resistance $R(T)$,
known as the Kondo effect. They concluded from those data that magnetic impurities were not the cause of the saturation
of $\tau_{\phi}$ they observed in their nominally pure gold samples. However, it is well known that the spin-flip
scattering rate peaks near the Kondo temperature $T_{K}$, then decreases at lower temperature. While MJW showed
convincingly that ``saturation'' of $\tau_{\phi}$ in gold could not be caused by iron impurities with
$T_{K}\approx0.3$~K, their data do not exclude an effect of impurities with a lower Kondo temperature, such as
manganese or chromium (see Table~\ref{tableMagImp}).

\begin{table}[ptbh]
\begin{tabular}
[c]{|c|ccc|}\hline
$_{\mathrm{Host}} \diagdown^{\mathrm{Impurity}}$~ & ~Cr~ & ~Fe~ & ~Mn~\\\hline
Ag & ~$\sim0.02$~ & ~$\sim3$~ & ~0.04~\\
Au & ~$\sim0.01$~ & ~0.3~ & ~$<0.01$~\\
Cu & ~1.0~ & ~25~ & ~0.01~\\\hline
\end{tabular}
\caption{Kondo temperature $T_{K}$ (K) of common, low $T_{K}$, magnetic impurities in Ag, Au and Cu (taken from
\cite{Wohlleben}).} \label{tableMagImp}
\end{table}

\subsection{Can dilute magnetic impurities account for a plateau in $\tau_{\phi}(T)$?}

To answer this question experimentally, we fabricated simultaneously three silver samples Ag(5N)b,
Ag(5N)c$_{\mathrm{Mn}0.3}$ and Ag(5N)d$_{\mathrm{Mn} 1}$, and very dilute manganese atoms were introduced by ion
implantation\cite{noteIonImp} in two of them. Manganese atoms form Kondo impurities in silver with a Kondo temperature
$T_{K}\simeq40$~mK.

\begin{figure}[ptbh]
\includegraphics[width=3.1in]{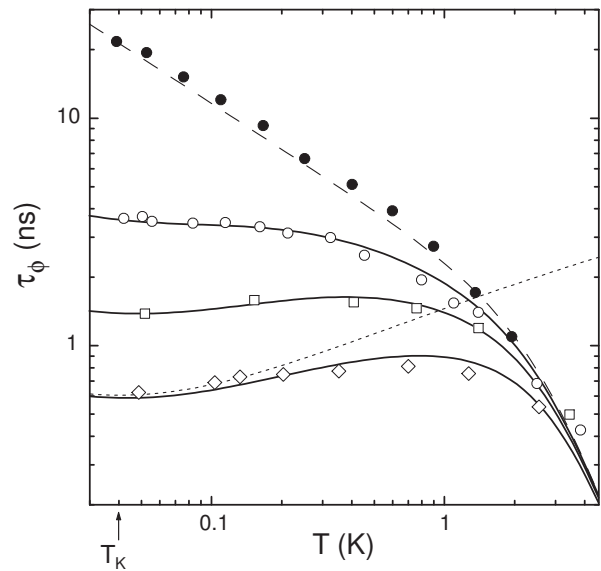} \caption{Phase coherence time as
function of temperature in several silver wires. Sample Ag(6N)c ($\bullet$) is made of the purest silver source.
Samples Ag(5N)b ($\circ$), Ag(5N)c$_{\mathrm{Mn}0.3}$ ($\square$) and Ag(5N)d$_{\mathrm{Mn}1}$ ($\diamond$) were
evaporated simultaneously using our 5N silver source. Afterward, 0.3~ppm and 1~ppm of manganese was added by ion
implantation respectively in samples Ag(5N)c$_{\mathrm{Mn}0.3}$ and Ag(5N)d$_{\mathrm{Mn}1}$. The presence of very
dilute manganese atoms, a magnetic impurity of Kondo temperature $T_{K}=40$~mK, reduces $\tau_{\phi}$ leading to an
apparent ``saturation'' at low temperature. Continuous lines are fits of $\tau_{\phi}(T)$ taking into account the
contributions of electron-electron and electron-phonon interactions (dashed line) and spin flip collisions using the
concentration $c_{\mathrm{mag}}$ of magnetic impurity as a fit parameter (dotted line is $\tau_{\mathrm{sf}}$ for
$c_{\mathrm{mag}}=1$~ppm). Best fits are obtained using $c_{\mathrm{mag}}=0.13,$ 0.39 and 0.96~ppm respectively for
samples Ag(5N)b, Ag(5N)c$_{\mathrm{Mn}0.3}$ and Ag(5N)d$_{\mathrm{Mn} 1}$, in close agreement with the concentrations
implanted and consistent with the source material purity used.} \label{FigAgMn}
\end{figure}

The phase coherence times extracted from WL corrections are shown as symbols in Fig.~\ref{FigAgMn}. Samples Ag(6N)c,
evaporated separately, is shown as a reference. At the time of this experiment only the 5N purity silver source was
available. Sample Ag(5N)b, in which no manganese atoms were implanted, already shows very little temperature dependence
of $\tau_{\phi }\sim3.5$~ns below 0.3~K. Nevertheless, introducing more manganese reduces further the phase coherence
time as illustrated with samples Ag(5N)c$_{\mathrm{Mn}0.3}$ and Ag(5N)d$_{\mathrm{Mn}1}$ in which respectively 0.3 and
1~ppm of manganese were implanted. For instance, by adding 1~ppm of manganese, $\tau_{\phi}$ was reduced by a factor of
6 while leaving $\tau_{\phi}$ still nearly independent of temperature.

The effect of manganese on $\tau_{\phi}$ is now compared with the existing theory of spin-flip scattering in the Kondo
regime.

\subsection{Comparison with the theory of spin-flip scattering}

In the presence of spin-flip scattering the phase coherence time reads
\begin{equation}
\frac{1}{\tau_{\phi}}=\frac{1}{\tau_{ee}}+\frac{1}{\tau_{\mathrm{ph}}}+\frac{1}{\tau_{\mathrm{sf}}},\label{fitee-eph-sf}
\end{equation}
where $1/\tau_{\mathrm{sf}}$ is the spin-flip rate of electrons. This
expression is valid when the spin-flip scattering time of the conduction
electrons is longer than the spin relaxation time ($\tau_{K}$ for Korringa
time) of the magnetic impurities themselves, \emph{i.e.} $\tau_{\mathrm{sf}%
}>\tau_{K}$.\cite{Falko} This holds if
\begin{equation}
T\gtrsim\frac{c_{\mathrm{mag}}}{\nu_{F}k_{B}}
\end{equation}
where $c_{\mathrm{mag}}$ is the concentration per unit volume of magnetic impurities. In silver, gold and copper
this criterion reads
\begin{equation}
T\gtrsim40~\mathrm{mK}\times~c_{\mathrm{mag}}[\mathrm{ppm}],
\end{equation}
in which $c_{\mathrm{mag}}[\mathrm{ppm}]$ is now written in parts per million atoms (ppm). In the opposite limit
($\tau_{\mathrm{sf}}<\tau_{K}$), the impact of spin flip scattering on $\tau_{\phi}$ depends on the physical effect
probed. For weak localization corrections with strong spin-orbit coupling, spin-flip scattering enters then as
$2/\tau_{\mathrm{sf}}$ in Eq.~(\ref{fitee-eph-sf}).\cite{dephasingMI,Falko}

As long as $T\gtrsim T_{K}$, $\tau_{\mathrm{sf}}$ is well described by the
Nagaoka-Suhl formula \cite{Maple,Haesendonck_NS}
\begin{equation}
\frac{1}{\tau_{\mathrm{sf}}}=\frac{c_{\mathrm{mag}}}{\pi\hbar\nu_{F}}
\frac{\pi^{2}S(S+1)}{\pi^{2}S(S+1)+\ln^{2}(T/T_{K})},\label{NagaokaSuhl}
\end{equation}
with $S$ and $T_{K}$ respectively the spin and Kondo temperature of the
magnetic impurities.

Upon cooling down, $\tau_{\mathrm{sf}}$ decreases when $T$ approaches $T_{K}$ (dotted line in Fig.~\ref{FigAgMn}),
whereas the electron-electron scattering time $\tau_{ee}$ increases. The combination of both contributions can result
in a nearly constant phase coherence time above $T_{K},$ as shown by the solid lines in Fig.~\ref{FigAgMn}.

A quick way to estimate the concentration of magnetic impurities corresponding to a plateau in the phase coherence time
is to compare $\tau_{\phi}^{\mathrm{plateau}}$ at the plateau to the prediction of Nagaoka-Suhl at $T=T_{K}$. In
samples made of copper, gold and silver this gives
\begin{equation}
\tau_{\phi}^{\mathrm{plateau}}~[\mathrm{ns}]\simeq0.6/c_{\mathrm{mag}}~[\mathrm{ppm}].
\end{equation}

\begin{figure}[ptbh]
\includegraphics[width=3.1in]{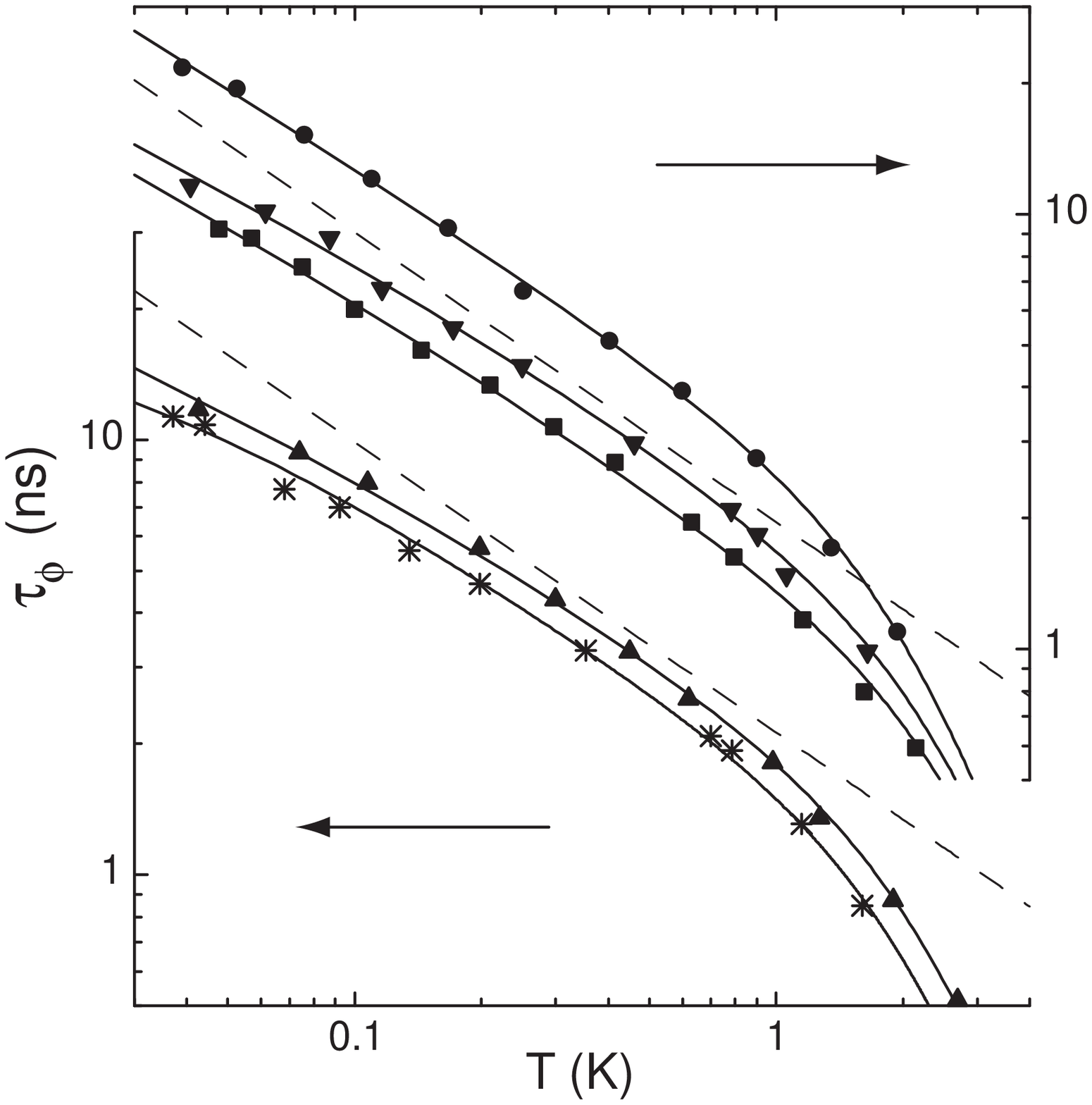}
\caption{Phase coherence time vs temperature measured on samples Ag(6N)a ($\blacksquare$), Ag(6N)b
($\blacktriangledown$), Ag(6N)c ($\bullet$), Ag(6N)d ($\blacktriangle$), and Au(6N) ($\ast$). For clarity the graph has
been split in two parts shifted vertically, as was done in Fig.~\ref{FigAg6N}. In contrast to Fig.~\ref{FigAg6N},
continuous lines are fits of the data to Eqs.~(\ref{fitee-eph-sf}) and (\ref{NagaokaSuhl}), with the concentration of
magnetic impurities as an additional fit parameter (see Table~\ref{tableFitMI}). The quantitative prediction of
Eq.~(\ref{tauEEtext}) for electron-electron interactions in samples Ag(6N)b (top part) and Ag(6N)d (bottom part) are
shown as dashed lines.} \label{Fig6NfitMI}
\end{figure}

\begin{table}[ptbh]
\begin{tabular}
[c]{|c|c|c|c|c|}\hline
Sample & $A~(A_{thy})$ & $B$ & $c_{\mathrm{mag}}$ & $T_K$\\
& (ns$^{-1}$K$^{-2/3}$) & (ns$^{-1}$K$^{-3}$) & (ppm) & (K) \\\hline
Ag(6N)a & 0.68 (0.55) & 0.051 & 0.009 & 0.04\\
Ag(6N)b & 0.54 (0.51) & 0.05 & 0.011 & 0.04\\
Ag(6N)c & 0.35 (0.31) & 0.051 & 0.0024 & 0.04\\
Ag(6N)d & 0.50 (0.47) & 0.054 & 0.012 & 0.04\\
Au(6N) & 0.59 (0.40) & 0.08 & 0.02 & 0.01\\\hline
\end{tabular}
\caption{Fit parameters for $\tau_{\phi}(T)$ in silver and gold samples made of our 6N sources, taking into account, on
top of the contributions of electron-electron and electron-phonon interactions, the additional contribution of dilute
Kondo impurities of spin $1/2$ as described by Eqs.~(\ref{fitee-eph-sf}) and (\ref{NagaokaSuhl}). The corresponding
fits are displayed as continuous lines in Fig.~\ref{Fig6NfitMI}.} \label{tableFitMI}
\end{table}

Continuous lines in Fig.~\ref{FigAgMn} are fits of the measured $\tau_{\phi }(T)$ to Eq.~(\ref{fitee-eph-sf}) using
Eq.~(\ref{NagaokaSuhl}), with magnetic impurities of Kondo temperature $T_{K}=40$~mK as expected for manganese atoms.
The parameters $A$ and $B$ in Eq.~(\ref{fitee-eph}) could not be extracted independently for samples Ag(5N)b,
c$_{\mathrm{Mn}0.3}$ \& d$_{\mathrm{Mn}1}$. To avoid increasing unnecessarily the number of fit parameters, the values
of $A$ and $B$ deduced from the fit of sample Ag(6N)c (dashed line) were used. Sample Ag(6N)c was chosen as a reference
because its predicted electron-electron scattering rate is close to that of samples Ag(5N)b, Ag(5N)c$_{\mathrm{Mn}0.3}$
and Ag(5N)d$_{\mathrm{Mn}1}$. Following this procedure, the measurements could be reproduced accurately
with\cite{noteSpinFit} $S=1/2$ and $c_{\mathrm{mag}}=0.13,$ 0.39 and 0.96~ppm, respectively for samples Ag(5N)b,
c$_{\mathrm{Mn}0.3}$ \& d$_{\mathrm{Mn}1}$, in close agreement with implanted concentrations of manganese and
compatible with the nominal purity of the Saclay 5N silver source. This confirms that the effect on $\tau_{\phi}$ of
the implantation of magnetic impurities with a low Kondo temperature is well understood, both qualitatively and
quantitatively.

Looking back at the $\tau_\phi$ data for samples Ag(6N)a, b, c, d and Au(6N) shown in Fig.~\ref{FigAg6N}, we note that
the fits to those data would also improve with the addition of a very small quantity of magnetic impurities. We
performed new fits to those data using Eqs.~(\ref{fitee-eph-sf}) and (\ref{NagaokaSuhl}), with $c_{\mathrm{mag}}$ as an
additional adjustable parameter. For the silver samples we kept $T_K = 40$~mK as for manganese impurity atoms, whereas
for the gold sample Au(6N) we chose $T_K = 10$~mK as for chromium impurity atoms. The values of $c_{\mathrm{mag}}$ from
the fits are 0.009, 0.011, 0.0024, 0.012, and 0.02~ppm respectively for samples Ag(6N)a, b, c, d, and Au(6N). The new
fits are shown as continuous lines in Fig.~\ref{Fig6NfitMI} and the fit parameters are given in Table~\ref{tableFitMI}.
Note that these concentrations are about 100 times smaller than the nominal total impurity concentrations of the
sources. As a striking example to show how small these numbers are, 0.01~ppm of impurities in sample Ag(6N)d
corresponds to about 3 impurity atoms every micrometer in the wire. Such small concentrations of Kondo impurities are
essentially undetectable by any means other than measuring the phase coherence time, especially in thin films.
Moreover, no commercial provider can guarantee such a high purity for the source material.

\subsection{Extremely dilute magnetic impurities and temperature dependence of the resistance}

The temperature dependence of the resistance, $R(T),$ is often used as a probe of magnetic impurities, because of the
well-known Kondo effect. Nevertheless, in thin wires, where the resistance also varies due to electron-electron
interactions, it must be pointed out that $R(T)$ is not sensitive enough to detect small amounts of magnetic
impurities. The contribution of electron-electron interactions\cite{AleinerWav}
\begin{equation}
\frac{\Delta R(T)}{R}\simeq3.126\frac{R}{R_{K}}\frac{L_{T}}{L} \equiv \frac{C_{\mathrm{thy}}}{\sqrt{T}}, \label{eqR(T)}
\end{equation}
with $L_T=\sqrt{\hbar D/k_B T}$ the thermal length, is much stronger and varies much more rapidly with
temperature than the Kondo term, determined by $\Delta\rho_{\mathrm{Kondo}}\simeq-B_{K} \ln (T)$,\cite{rhoKondo}
where $B_{K}\approx0.2$~n$\Omega$.cm/ppm.\cite{rhoKondoB} In our wires where the resistivity is about
$\rho\sim3~\mu\Omega$.cm, the corresponding relative variation of the resistance is about $10^{-5}$ per decade
of temperature for 1~ppm of Kondo impurities. This is more than an order of magnitude smaller than the typical
contribution of electron-electron interactions between 100~mK and 1~K.

\begin{figure}[ptbh]
\includegraphics[width=3.4in]{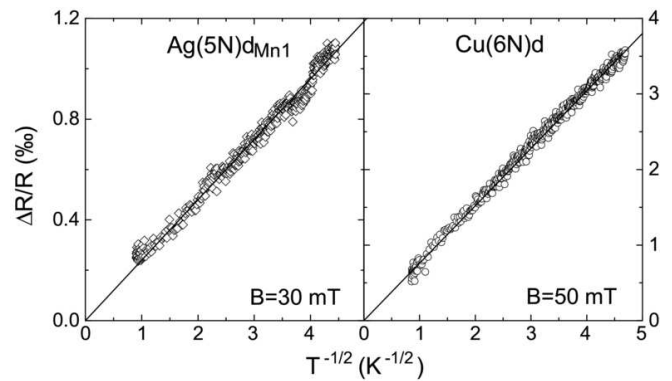} \caption{Resistance of sample
Ag(5N)d$_{\mathrm{Mn}1}$ ($\diamond$) and Cu(6N)d ($\circ$) plotted as function of $1/\sqrt{T}$. Continuous lines are
fits using the functional form $\Delta R(T) / R  = C/\sqrt{T},$ with $C=2.4\,10^{-4}$ (left panel) and
$7.6\,10^{-4}~\mathrm{K}^{1/2}$ (right panel), close to the predictions of Eq.~(\ref{eqR(T)})
$C_{\mathrm{thy}}=1.8\,10^{-4}$ and $7.2\,10^{-4}~\mathrm{K}^{1/2}$, respectively. The logarithmic contribution to
$R(T)$ from the Kondo effect is invisible in both samples, as it is masked by the much larger contribution from
electron-electron interactions in the wires. From the comparison of Figs.~\ref{FigAgMn} and \ref{FigRofT}, it appears
clearly that the phase coherence time is a much more sensitive probe of very dilute magnetic impurities than the
temperature dependence of the resistance.} \label{FigRofT}
\end{figure}

This is illustrated in the left panel of Fig.~\ref{FigRofT} with sample Ag(5N)d$_{\mathrm{Mn}1}$ in which we implanted
1~ppm of manganese. The resistances are measured in a magnetic field $B\sim20-50$~mT, large enough to suppress the WL
corrections but small enough to avoid freezing out the spin-flip scattering of conduction electrons by magnetic
impurities. We checked on several samples showing anomalous dephasing that $R(T)$ is independent of the applied
magnetic field.

A striking conclusion is that the phase coherence time is a much more
sensitive probe of very dilute magnetic impurities than the temperature
dependence of the resistance, which is dominated by electron-electron
interactions at low temperature.

\section{Other explanations of anomalous dephasing}

The evidence presented in the previous section shows that very dilute magnetic impurities could explain the anomalous
dephasing frequently observed at low temperature. But are there other viable explanations?

\subsection{Dephasing by high energy electromagnetic modes}

Golubev and Zaikin (GZ) proposed\cite{GZ,GZS} that zero temperature dephasing by high energy electromagnetic modes is
responsible for the frequently observed saturation of $\tau_{\phi}$ in metallic thin films. This theory, which is
controversial,\cite{controversyGZS} predicts that the phase coherence time saturates at low temperature at
$\tau_{0}^{\mathrm{GZ}}$ given by\cite{GZS}
\begin{equation}
\frac{1}{\tau_{0}^{\mathrm{GZ}}}=\frac{\sqrt{2}\rho}{3R_{K}\pi\sqrt{D}}\left( \frac{b}{\tau_{e}}\right)
^{3/2},\label{tau0}
\end{equation}
where $b$ is a constant numerical factor expected to be of order 1. It is
interesting to point out that for a given material $\tau_{0}^{\mathrm{GZ}}$ is
proportional to $D^{3}$ and is insensitive to the actual geometry of the sample.

Using this prediction, GZ were able to account for a subset of the experimental results published in
\cite{Gougam,Natelson} using the overall prefactor of the dephasing rate as an adjustable parameter.\cite{GZS}
Note that, as explained by GZ in their latest article,\cite{GZS} the comparison with MJW data performed in
\cite{GZ_PRB} should be ignored because it was done using an expression for $\tau_{0}^{\mathrm{GZ}}$ that does
not apply to the experiment, but is valid only when the elastic mean free path exceeds the transverse dimensions
of the wires.

\begin{figure}[ptbh]
\includegraphics[width=3.4in]{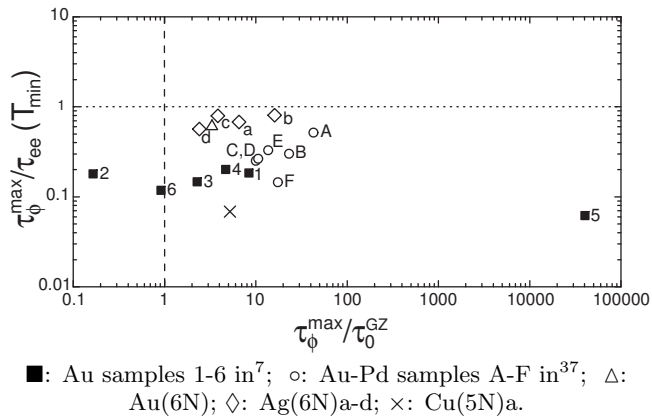}  \newline $\blacksquare$: Au
samples 1-6 in \cite{MJW};~ $\circ$: Au-Pd samples A-F in \cite{Natelson};~ $\vartriangle$: Au(6N); $\lozenge$:
Ag(6N)a-d; $\times$: Cu(5N)a.\caption{Comparison between the predictive powers of the conventional theory of
electron-electron interactions,\cite{AAK} and of the theory of Golubev and Zaikin.\cite{GZ,GZS} The X coordinate gives
the ratio of the phase coherence time measured at the lowest temperature, $\tau_{\phi }^{\mathrm{max}}$, to
$\tau_{0}^{\mathrm{GZ}}$, calculated from Eq.~(\ref{tau0}) with $b=1$. The Y coordinate is the ratio of
$\tau_{\phi}^{\mathrm{max}}$ to $\tau_{ee}(T_{\mathrm{min}} )$, the value calculated using the conventional theory
(Eq.~(\ref{tauEEtext})) at the lowest temperature $T_{\mathrm{min}}.$ Open symbols are data points for which the phase
coherence time continues to increase at the lowest measurement temperature. Full symbols and $\times$ are data points
for which the phase coherence time is nearly constant at low temperature. The conventional theory predicts that all
data points lie on the horizontal dotted line if no extrinsic degrees of freedom, such as magnetic impurities, limit
the phase coherence time. The GZ theory predicts that all data points lie on a vertical line if the phase coherence
time already saturates, and to the left of that line if $\tau_{\phi}$ still increases at low temperature. (The dashed
line corresponds to the case $b=1$ in the GZ theory).} \label{FigZaikin}
\end{figure}

Since the exact prefactor is unknown, it is not possible to rule out this theory by comparison with a single
experiment. Instead, we propose here to compare the predictive power of the GZ theory with the conventional theory of
electron-electron interactions for many samples. This is done in Fig.~\ref{FigZaikin}. This figure includes all gold,
silver and gold-palladium samples for which it has not been shown that magnetic impurities are the main source of
decoherence at low temperature, plus sample Cu(5N)a which was used by GZ for comparison of their theory with
experiments.\cite{GZS} (We do not show other copper samples or samples made from our 5N silver source, because they
clearly contain magnetic impurities. See section~VII and \cite{Anthore}.) The X coordinate in Fig.~\ref{FigZaikin}
gives the ratio of the phase coherence time measured at the lowest temperature, $\tau_{\phi}^{\mathrm{max}}$, to
$\tau_{0}^{\mathrm{GZ}}$, calculated from Eq.~(\ref{tau0}) with $b=1$. The Y coordinate is the ratio of
$\tau_{\phi}^{\mathrm{max}}$ to $\tau_{ee}(T_{\mathrm{min}})$, the value calculated using the conventional theory
(Eq.~(\ref{tauEEtext})) at the lowest temperature $T_{\mathrm{min}}$. Open symbols are samples for which $\tau _{\phi}$
continues to increase at the lowest measurement temperature; upon cooling they move to the right. Full symbols are
samples for which $\tau_{\phi}$ is nearly constant at low temperature; they move downward when the temperature is
reduced. As for theory, GZ predict that all full symbols should be aligned on a vertical line
$\tau_{\phi}^{\mathrm{max} }/\tau_{0}^{\mathrm{GZ}}=b^{3/2}$, whereas open symbols would be located at
$\tau_{\phi}^{\mathrm{max}}/\tau_{0}^{\mathrm{GZ}}<b^{3/2}$. In contrast, the conventional theory predicts that all
data points should be aligned on the horizontal line $\tau_{\phi}^{\mathrm{max}}/\tau_{ee}(T_{\mathrm{min} })=1$. On
this plot the data scatter in both directions. The most salient feature of the plot, however, is that the scatter in
the horizontal direction extends over more than 5 orders of magnitude, whereas the scatter in the vertical direction
extends over slightly more than one decade. The horizontal scatter indicates that GZ theory does not reproduce the
dependence of $\tau_{\phi}$ on sample parameters. In particular, the GZ prediction depends much too strongly on the
diffusion coefficient, which varies considerably in MJW's six gold samples.

While no theory explains all of the experimental data without any additional parameters, it appears that the
conventional theory does a better job than the GZ theory to predict the low temperature value of $\tau_{\phi}$.

\subsection{Dephasing by two level systems}

Two approaches to electron dephasing by two-level tunneling systems (TLS) have been proposed. The first, by Imry,
Fukuyama, and Schwab,\cite{IFS} requires a non-standard distribution of TLS parameters. It was shown later that such a
distribution would lead to large anomalies in the low-temperature specific heat, and in acoustic attenuation at very
low temperature.\cite{AleinerTLS} The second approach describes the coupling between the conduction electrons and the
TLS via the two-channel Kondo effect.\cite{2CK} In this model, the effect of TLS is very similar to that of magnetic
impurities in the Kondo regime, at least at $T\gtrsim T_{K}$. The main criticism raised against this explanation is
that, starting from any realistic model of a TLS, it may be impossible to reach the strong coupling regime where the
Kondo temperature is larger than the tunneling level splitting.\cite{TLS_TK_Aleiner,TLS_TK_Zawa} From the experimental
point of view, measurements of $\tau_{\phi}$ from the weak localization contribution to the magnetoresistance cannot
discriminate between magnetic impurities and TLS.

\section{Test of the magnetic impurity hypothesis: probing $\tau_{\phi}(B)$}

A definitive test of the role of spin-flip scattering for the saturation of $\tau_{\phi}$ at low temperature is to
probe how the dephasing time depends on magnetic field. It is expected that spin-flip scattering is suppressed when the
dynamics of magnetic impurities is frozen by application of a sufficiently large magnetic field $B$. Indeed, if the
Zeeman splitting is much larger than $k_{B}T$, magnetic impurities stay in their ground state. As a result spin-flip
collisions vanish and $\tau_{\phi}$ should climb up to the value expected from electron-electron interactions
(independent of $B$ as long as the cyclotron radius is much larger than the elastic mean free path). In the presence of
spin $1/2$ impurities, and neglecting Kondo effect, the spin-flip scattering rate of electrons vanishes at large field
as (see Appendix~C and \cite{Meyer})
\begin{equation}
\frac{\tau_{\mathrm{sf}}(B=0)}{\tau_{\mathrm{sf}}(B)}=\frac{g\mu B/k_{B} T}{\sinh(g\mu B/k_{B}T)},\label{tausfB}
\end{equation}
where $g$ is the renormalized gyromagnetic factor of the magnetic impurities.

One possible method to detect a variation in $\tau_{\phi}$ with magnetic field
is to measure the average amplitude $\Delta G_{\mathrm{UCF}}$ of universal
conductance fluctuations in a metallic wire as a function of magnetic field.
This method has two drawbacks. First $\Delta G_{\mathrm{UCF}}\propto\tau
_{\phi}^{1/4}$ depends only weakly on the phase coherence time. Second the
large correlation field $\Delta B_{\mathrm{UCF} }\simeq h/(ewL_{\phi})$ of
conductance fluctuations in mesoscopic wires makes it difficult to obtain
accurate estimates of the averaged $\Delta G_{\mathrm{UCF}}(B)$ at low
temperature in the field range below the relevant magnetic field scale $g\mu B\sim
k_{B}T$. For example, in Cu(6N)b, $\Delta B_{\mathrm{UCF}} \simeq
25~\mathrm{mT}$ at 40~mK, whereas the characteristic field needed to freeze
the magnetic impurities is as low as $k_{B}T/2\mu\simeq55~\mathrm{mT.}$

We have chosen instead to probe the magnetic field dependence of $\tau_{\phi}$ by measuring the Aharonov-Bohm (AB)
oscillations in the magnetoresistance of ring-shaped samples. For this purpose, we have fabricated two copper rings of
radius $r=0.5$ and 0.75~$\mu$m respectively on the same chip as samples Cu(6N)c and Cu(6N)d. The ring perimeters are
chosen to be larger than or similar to the phase coherence length at $B\approx0$ in order to increase the sensitivity
to variations of $\tau_{\phi}$. The averaged $h/e$ AB oscillations amplitude $\Delta G_{\mathrm{AB}}$ is related to the
phase coherence time through\cite{NoteAB}
\begin{equation}
\Delta G_{\mathrm{AB}}=C\frac{e^{2}}{h}\frac{L_{T}}{\pi r}\sqrt{\frac{L_{\phi
}}{\pi r}}\exp\left[  -\frac{\pi r}{L_{\phi}}\right]  ,\label{GAB}%
\end{equation}
where $C$ is a geometrical factor of order 1. The short period of AB oscillations with $B$ (5.5 and 2.5~mT for $r=0.5$
and 0.75~$\mu$m, respectively) allows to estimate accurately the magnetic field dependence of $\Delta G_{\mathrm{AB}}$
on the much larger field scale needed to freeze the magnetic impurities.

This experiment was performed on copper samples because it is the material in which the presence of magnetic
impurity was most questionable: no correlations were found between $\tau_{\phi}$ and the copper source material
purity; moreover, whereas in some samples $\tau_{\phi}$ saturates at values as small as 0.2~ns (3 times smaller
than in Ag(5N)d$_{\mathrm{Mn}1}$) we observed neither a non-monotonic temperature dependence of
$\tau_{\phi}(T)$, as in Ag(5N)d$_{\mathrm{Mn}1}$ (see Fig.~\ref{FigAgMn}), nor a Kondo contribution to $R(T)$.

\begin{figure}[ptbh]
\includegraphics[width=3.2in]{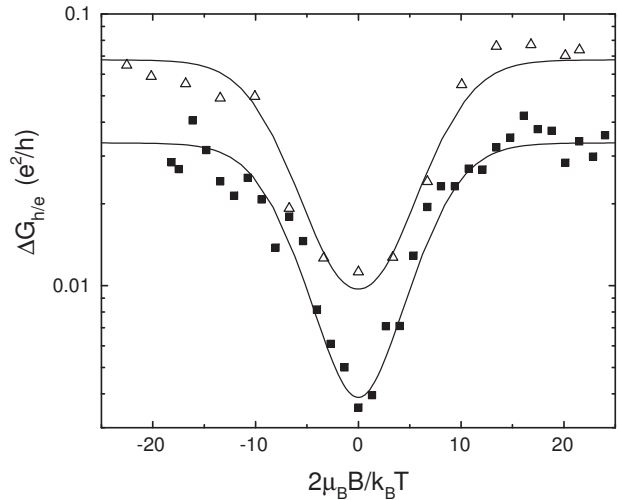} \caption{Symbols: mean amplitude of
the AB $h/e$ oscillations ($\Delta G_{h/e}$) across the ring in sample Cu(6N)d at $T=40$ ($\vartriangle$) and 100~mK
($\blacksquare$), plotted in units of $e^{2}/h$ as a function of the reduced magnetic field $2\mu_{B}B/k_{B}T$. Solid
lines: fits to the two data sets using Eqs.~(\ref{fitee-eph-sf}), (\ref{tausfB}) and (\ref{GAB}) with $C$ and $g$ as
fit parameters. At 40~mK, the AB oscillations are unmeasurably small at $B=0$; the fit to those data includes the noise
floor of the experiment.} \label{FigAB}
\end{figure}

Our experimental procedure and data analysis are detailed in \cite{PierreAB}. Figure~\ref{FigAB} shows the amplitude of
AB oscillations measured across the ring in sample Cu(6N)d at $T=40$ and $100$~mK (symbols) as a function of reduced
magnetic field $2\mu B/k_{B}T$. The data in Fig.~\ref{FigAB} show that the amplitude of AB oscillations increases with
magnetic field by a factor 8 at 100~mK and a factor 7 at 40~mK,\cite{NoteABnoise} on a characteristic field scale
proportional to $T$.

The solid lines in Fig.~\ref{FigAB} are fits to the simple model represented by Eqs.~(\ref{tausfB}) and
(\ref{GAB}), explained in Appendix~C. We assumed that $\tau_{\phi}$ at large $B$ is limited only by
electron-electron interactions and used the values given by theoretical prediction (Eq.~(\ref{tauEEtext})):
$\tau_{\phi}=5.4$ and $9.9~\mathrm{ns}$ at 100 and 40~mK, respectively. The two remaining parameters, namely the
gyromagnetic factor $g$ and the geometrical constant\cite{ABfitnote} $C$, were adjusted to reproduce accurately
our data. The best fit is obtained with $g=1.08$ and $C=0.17$. Note that a more rigorous approach to the
magnetic-field dependence of AB oscillation amplitude has been published recently by Vavilov and
Glazman.\cite{VavilovGlazman} Using their prediction (Eqs.~(62) and (63) in \cite{VavilovGlazman}) with a
magnetic impurity spin\cite{VG_S} $S=1/2$ and $g=0.90$, we obtain a fit indistinguishable from the solid lines
calculated with the simple model.

The impurity $g$-factors obtained from these fits, 1.08 and 0.90, are small, like the value $g=1.36$ found for
electrons by neutron scattering in bulk CuO.\cite{Shirane}

This set of experiments confirms that spin-flip collisions are responsible for
the apparent low temperature saturation of $\tau_{\phi}$ we observe in copper samples.

\section{Comparison with energy exchange measurements}

Parallel to this work, a systematic correlation was found between dephasing and energy exchange between electrons: all
samples made of the same source material, using the same deposition system, either followed the theory of
electron-electron interactions for both energy exchange and phase coherence, or displayed anomalous behaviors for both
phenomena.\cite{Gougam,PierreJLTP,PierrePesc,PierrePHD} This correlation suggests that magnetic impurities could also
be responsible for anomalous energy exchange. Such a possibility had not been considered until recently because, all
spin states being degenerate at zero magnetic field, magnetic impurities do not contribute to energy exchange in first
order. However, Kaminsky and Glazman have pointed out that energy exchange in the presence of magnetic impurities may
take place with an appreciable efficiency by a second-order process.\cite{KaminskyGlazman} The experimental proof that
excess energy exchange observed in samples made of the 5N silver and copper sources result from dilute paramagnetic
spins was obtained recently by measuring the dependence of energy exchange upon magnetic field.\cite{Anthore} Similarly
to what was observed on the dephasing rate, the application of a large magnetic field on these samples reduces the rate
of energy exchange. Note however that the amount of magnetic impurities needed to account for the measured energy
exchange rates seems to be significantly larger than the estimations from $\tau_{\phi}(T)$; in the case of copper, the
obtained g-factor $g=2.3$ is also different. More experiments are needed to clarify these issues.

\section{Conclusion}

By measuring the phase coherence time as a function of temperature on wires made of silver, gold and copper,
from source materials of different purities, we have found that anomalous dephasing is correlated to source
material purity in silver and gold samples, and systematic in copper samples. We showed experimentally that the
presence of very dilute magnetic impurities with a low Kondo temperature in the host material can result in a
broad plateau in $\tau_{\phi}(T)$ while being undetected in the temperature dependence of the resistance.
Measurement of the magnetic field dependence of Aharonov-Bohm oscillations on relatively large copper rings
revealed that the phase coherence time increases with $B$ on a field scale proportional to the temperature. This
confirms that an apparent ``saturation'' of $\tau_{\phi}$ can be attributed to very dilute magnetic
impurities.\cite{JJLin}

In the silver and gold samples discussed in this paper, we impute the presence of magnetic impurities to the
purity of the material sources. We found that large coherence times at 40~mK could be obtained in samples
fabricated with the silver sources of the highest purity commercially available (6N). However, it is very
difficult to rule out a small contamination during the evaporation process and eventually sample preparation. In
the case of copper, the Kondo impurities probably originate from the copper oxide at the
surface.\cite{Haesendonck}

This work was supported by NSF grants DMR-9801841 and 0104178, and by the Keck Microfabrication Facility supported by
NSF DMR-9809688. We acknowledge the assistance of S.~Gautrot, O.~Kaitasov, and J.~Chaumont at the CSNSM in Orsay
University, who performed the ion implantation in samples Ag(5N)c$_{\mathrm{Mn}0.3}$ and Ag(5N)d$_{\mathrm{Mn}1}$. We
are grateful to I.~Aleiner, B.L.~Altshuler, H.~Bouchiat, M.H.~Devoret, V.I.~Fal'ko, L.I.~Glazman, D.~Natelson,
M.G.~Vavilov and A.D.~Zaikin for interesting discussions.

\section{Appendix A: Electron cooling in transport measurements at low temperatures}

\begin{figure}[ptbh]
\includegraphics[width=3.4in]{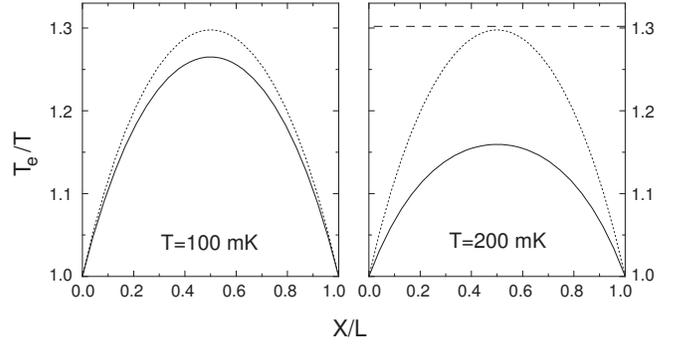} \caption{Electrons heating in a
typical silver wire (see text) of length $L=0.2$~mm, biased with a DC voltage $V$ such that $eV/k_{B}T=3$ and for
phonon temperatures $T=100$ and $200$~mK respectively in the left and right panel. Continuous lines: ratio of electron
temperature $T_{e}$ with phonon temperature as function of the reduced position $X/L$ in the wire, taking into account
electron-phonon interactions (see Eq.~(\ref{Heating})). Dotted lines: electron temperature as function of position
neglecting phonons (see Eq.~(\ref{HotEregime})). Dashed line in right panel: electron temperature neglecting electronic
heat transport (in the left panel this line would stand at $T_e/T=1.87$).} \label{FigHeating}
\end{figure}

Joule heating is a concern when transport measurements are performed at low temperatures. Any current results in the
production of heat, which can be either transferred directly to the phonons in the wire, or to the electrons in the
contact pads, assumed to be much larger than the wire. At sub-Kelvin temperatures, the first process becomes very
inefficient. The reason is that the phonon emission rate for an electron with an excess energy $k_{B}T$ goes
like\cite{PierrePHD} $\gamma \simeq 5 \kappa_{\mathrm{ph}}(k_{B}T)^{3},$ with
$\kappa_{\mathrm{ph}}\simeq10~\mathrm{ns}^{-1}\mathrm{meV}^{-3}.$ The distance it will travel before losing its extra
energy is then $\sqrt{D/\gamma} \simeq 18~\mu\mathrm{m\times}\left(  \frac{T}{1~\mathrm{K}}\right) ^{-3/2}$ for a
typical diffusion coefficient $D=100~\mathrm{cm}^{2}/\mathrm{s}.$ At $T=40~\mathrm{mK},$ $\sqrt{D/\gamma}\simeq
2.2~\mathrm{mm,}$ a very macroscopic distance! Therefore one must take care that the electron's energy never gets so
large at low temperature. Taken alone, the cooling by the contact pads through electronic heat transport results in a
temperature profile in the wire
\begin{equation}
T_{e}(x)=\sqrt{T^{2}+\frac{3}{\pi^{2}}x(1-x)\left(  \frac{eV}{k_{B}}\right) ^{2}},\label{HotEregime}
\end{equation}
with $T_{e}$ the electron temperature in the contacts placed at the ends of the wire, assumed to be equal to the
temperature of the phonons, $x$ the relative position along the wire, and $V$ the voltage across the wire. For $T=0$,
the maximum temperature is $\frac{\sqrt{3}}{2\pi}\left(  \frac{eV}{k_{B}}\right)
\approx3.2~\mathrm{K}\times\frac{V}{1~\mathrm{mV}}.$ By limiting the voltage across the sample to $eV=k_{B}T,$ the
maximal electron temperature is $T\sqrt{1+\frac{3}{4\pi^{2}}}\simeq1.04~T.$ With such a low applied voltage, the phase
coherence time, supposed to increase as $T_{e}^{-2/3}$ at low temperature, varies through the sample by $1-1.04^{-2/3}
\simeq 2\%,$ which is sufficiently small for most purposes. However, at very low temperature, a measurement of a
voltage of order $k_{B}T/e$ might become very time consuming if one considers that the input voltage noise for the best
room-temperature commercial amplifiers is about $1~\mathrm{nV}/\sqrt {\mathrm{Hz}}$ and that the weak localization
correction to the conductance is about $10^{-3}$ of the total signal. For example at 10~mK, $10^{-3}
k_{B}T/e\simeq1~\mathrm{nV},$ and an integration time of 100~s for each conductance measurement is needed to get a
signal to noise ratio of 10. In fact, this estimation is often too pessimistic because cooling by phonons does play a
role for long wires.\cite{HennySchonenberger} In order to evaluate this effect precisely, one has to solve the complete
heat equation, which can be written in reduced units ($t_{e}(x)=\frac{T_{e}(x)}{T},$ $v=\frac{eV}{k_{B}T}$)
\begin{equation}
v^{2}+\frac{\pi^{2}}{6}\frac{d^{2}}{dx^{2}}t_{e}^{2}(x)-\left(  \frac{T}{T_{co}}\right) ^{3}\left(
t_{e}^{5}(x)-1\right)  =0, \label{Heating}
\end{equation}
in which the first term describes Joule heating, the second the thermal conductivity of electrons, assuming
Wiedemann-Franz law, and the last one the coupling to phonons.\cite{Wellstood,PierrePHD} We have defined a
crossover temperature
\begin{equation}
T_{co}=(\Sigma\rho L^{2}(e/k_{B})^{2})^{-1/3}, \label{Tco}
\end{equation}
with $L$ the length of the wire, $\rho$ its resistivity, $\Sigma$ the electron-phonon coupling constant\cite{sigma}
(typically $\Sigma \sim1-10~\mathrm{nW}/\mu\mathrm{m}^{3}/\mathrm{K}^{5}$ in metallic thin films on Si substrate). The
resulting temperature profile is shown in Fig.~\ref{FigHeating} for typical values: we consider a silver wire
($\Sigma\simeq3~\mathrm{nW}/\mu\mathrm{m}^{3}/\mathrm{K}^{5}$ from Table~\ref{tableFitEE}) with
$D=100$~$\mathrm{cm}^{2}\mathrm{/s}$, $L=0.2$~$\mathrm{mm}$, at $T=200$~$\mathrm{mK}$, for $\frac{eV}{k_{B}T}=3.$ The
dotted line indicates the solution without phonons, the dashed line the solution without electronic heat transport. For
this set of parameters, the crossover temperature is $T_{co}\simeq120$~mK. Hence, at 200~mK phonons reduce
significantly the maximum electron temperature, which does not exceed the bath temperature by more than 16\%. At
100~mK, cooling by phonon emission is inefficient, and the maximum electron temperature is 27\% above $T.$

The analysis of the exact solutions of this equation allows to distinguish two opposite regimes: for $T\ll T_{co},$
electrons are decoupled from phonons (cooling by phonons will become active only if the applied voltage is so high that
the maximal temperature is above $T_{co}$), and temperature is given by the electronic conductivity alone, see
Eq.~(\ref{HotEregime}). This is the difficult regime, where the maximal voltage is of the order of $k_{B}T/e.$ In the
opposite situation $T\gg T_{co},$ heat transfer to the contacts can be neglected, and cooling by phonons rules the
game. The temperature of the electrons is then nearly homogeneous, with $\frac{T_{e}}{T}\approx\left( 1+\left(
\frac{T_{co}}{T}\right) ^{3}v^{2}\right) ^{1/5}$ and for $\left( \frac{T_{co}}{T}\right) ^{3}v^{2} \ll1$ the
temperature does not exceed $T$ excessively: $T_{e}\approx T+\frac{1}{5}\frac{T_{co}^{3}(eV/k_{B})^{2}}{T^{4} }.$ One
should thus fabricate wires as long as possible, in order to have a small crossover temperature $T_{co}$ which allows
to work at larger voltages.

\begin{figure}[ptbh]
\includegraphics[width=3.1in]{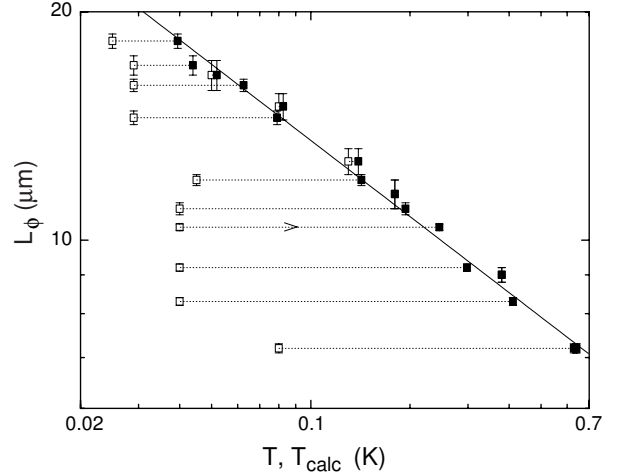} \caption{Full symbols: phase
coherence length measured on a 6N silver sample as a function of the electronic temperature $T_{\mathrm{calc}}$
calculated using Eq.~(\ref{Heating}) for a cryostat temperature $T$ represented by the attached open symbol. The
continuous line represents the theoretical prediction $L_{\phi}\varpropto T^{-1/3}$ of electron-electron interactions
(data taken at Saclay).} \label{FigHeatingExp}
\end{figure}

In order to test the validity of this calculation, we performed a control experiment in which electrons were
intentionally heated by applying \emph{ac} currents. The sample, similar to the others presented in this review,
consists of a $1.79$~$\mathrm{mm}$-long, 150 nm-wide and 45 nm-thick wire made out of a 6N purity silver source. The
diffusion coefficient $D=139$~$\mathrm{cm}^{2}/\mathrm{s}$ results in a cross-over temperature $T_{co}=30~\mathrm{mK.}$
We extracted the phase coherence length $L_{\phi}$ from the magnetoresistance. For each magnetoresistance trace we show
in Fig.~\ref{FigHeatingExp} two symbols, one open and one full, at a Y-coordinate given by the corresponding value of
$L_{\phi}$. Open symbols are at the X-coordinate given by the cryostat temperature $T$ at which the measurement was
performed, whereas full symbols are at the X-coordinate given by the calculated electron temperature
$T_{\mathrm{calc}}.$ Since the magnetoresistance is given by $L_{\phi} \varpropto T^{-1/3}$, $T_{\mathrm{calc}}$ was
calculated from the time- and position-average of $T^{-1/3}_e$, using temperature profiles obtained with
Eq.~(\ref{Heating}). For example, the pair of data points at $L_{\phi} \simeq10.4~\mathrm{\mu m}$ corresponds to
$T=40~\mathrm{mK,}$ $V_{ac}=0.86~\mathrm{mV}rms,$ leading to $T_{\mathrm{calc}}=245~\mathrm{mK}$. The data points with
large heating ($T_{\mathrm{calc}}\gg T$) as well as those with little heating ($T_{\mathrm{calc}}\simeq T$) fall close
to a single line $L_{\phi} \varpropto T^{-1/3}$, indicating that the electron temperature is correctly modelled.

\section{Appendix B: Dephasing by electron-electron interactions}

Assuming that we can restrict ourself to two body interactions, the dephasing
rate, or inverse lifetime, $1/\tau_{\mathrm{in}}(E,T)$ of an electron at
energy $E$ coupled only to the electronic fluid at temperature $T $ results
from all collision processes allowed by the Pauli exclusion principle:
\begin{equation}
\label{teegal}\tau_{\mathrm{in}}^{-1}(E,T) \simeq\int_{|\varepsilon|
\gtrsim\hbar/\tau_{\phi}}d\varepsilon~K(\varepsilon)(1-f_{T}(E-\varepsilon))
h(\varepsilon,T)
\end{equation}
where $f_{T}(E)$ is the Fermi function at temperature $T$, $K(\varepsilon)$ is
the interaction ``Kernel" of the screened Coulomb interaction, proportional to
the modulus square of the interaction matrix element for an exchanged energy
$\varepsilon$, and
\begin{eqnarray}
h(\varepsilon,T)  &=& \int_{-\infty}^{\infty} dE^{\prime}f_{T}(E^{\prime})
(1-f_{T}(E^{\prime}+\varepsilon))\nonumber\\
&=& \frac{\varepsilon}{1-\exp(-\varepsilon/k_{B}T)}.
\end{eqnarray}

The low energy cut-off $|\varepsilon| \gtrsim\hbar/\tau_{\phi}$ in Eq.~(\ref{teegal}) is introduced because
fluctuations on timescales longer than the electron's life-time can be considered as static.\cite{AA}

The interaction kernel $K(\varepsilon)$ depends only on $\varepsilon$ since
the energies of interacting electrons are close to the Fermi energy $E_{F}$
and $\varepsilon\lesssim k_{B}T \ll E_{F}$. Our samples are quasi-1D because
the width and thickness of the wires are smaller than the length
$L_{\varepsilon}=\sqrt{\hbar D/\varepsilon}$ for the probed energy exchanged.
For quasi-1D samples the interaction kernel reads\cite{AndreevKamenev}
\begin{equation}
\label{Kee}K(\varepsilon) = \kappa|\varepsilon|^{-3/2},
\end{equation}
with
\begin{equation}
\label{Kappa_ee}\kappa^{-1} = \hbar\sqrt{\frac{\pi\nu_{F} S L}{4}\frac{R_{K}%
}{R}}.
\end{equation}

The dephasing rate $1/\tau_{\mathrm{ee}}(T)$ is the inverse lifetime averaged
over thermal excitations
\begin{equation}
\label{tphiint}1/\tau_{\mathrm{ee}}(T)=\int dE \frac{f_{T}(E)(1-f_{T}%
(E))}{k_{B}T}\tau_{\mathrm{in}}^{-1}(E,T).
\end{equation}

Injecting Eqs.~(\ref{teegal}) and (\ref{Kee}) into Eq.~(\ref{tphiint}) we
obtain\cite{noteReplaceCutoff}
\begin{equation}
1/\tau_{\mathrm{ee}}(T)\simeq\int_{\hbar/\tau_{\mathrm{ee}}}^{\infty
}d\varepsilon\frac{\kappa\sqrt{\varepsilon}}{k_{B}T}\frac{\exp(\varepsilon
/k_{B}T)}{(1-\exp(\varepsilon/k_{B}T))^{2}}.
\end{equation}
This expression shows that the effect of electron-electron interactions on quantum coherence in mesoscopic wires is
dominated by processes with a small exchanged energy $\varepsilon\approx\hbar/\tau_{\phi}$. It is interesting to point
out that this implies that a sample is quasi-1D with respect to decoherence as long as the phase coherence length
$L_{\phi}=\sqrt{D\tau_{\phi}}$ is large compared to its transverse dimensions and small compared to its length. This is
not true for energy exchange, for which the dimensionality is determined by the length associated with the largest
exchanged energy.

In order to obtain an analytical expression for $\tau_{\mathrm{ee}}(T)$ we
make the following approximation:
\begin{equation}
\frac{\exp(\varepsilon/k_{B}T)}{(1-\exp(\varepsilon/k_{B}T))^{2}} \simeq\frac{1}{(\varepsilon/k_{B}T)^{2}}.
\end{equation}
This approximation is justified since the integral is dominated by small
energy exchanges. This leads to
\begin{equation}
\tau_{\mathrm{ee}}\simeq\hbar\left[  \frac{(\pi/16)(R_{\mathrm{K}}/R)\nu
_{F}SL}{(k_{B}T)^{2}}\right]  ^{1/3},\label{tauphiEE_Keps}%
\end{equation}
where we used Eq.~(\ref{Kappa_ee}) for the interaction kernel.

The calculation of $\tau_{\phi}$ described above makes use of a low energy cut-off, therefore the prefactor in
Eq.~(\ref{tauphiEE_Keps}) is not reliable. To solve this technical difficulty, Altshuler, Aronov and
Khmelnitsky\cite{AAK} calculated the effect of electron-electron interactions through the interaction of one electron
with the fluctuating electromagnetic field resulting from other electrons at thermal equilibrium. Within this approach
it is possible to calculate directly the conductivity taking into account electron-electron interactions. The dephasing
rate is then obtained without reference to the energy decay rate. Neglecting spin-orbit coupling, this calculation
yields\cite{AleinerWav}
\begin{equation}
\frac{\Delta R}{R}(B,T)=-\frac{2R}{R_{K}}\frac{\sqrt{D\tau_{N}}}%
{L}\frac{\mathrm{Ai}(\tau_{N}/\tau_{H})}{\mathrm{Ai^{\prime}}(\tau_{N} /\tau_{H})},\label{WLairy}
\end{equation}
with
\begin{eqnarray}
\tau_{N}  &=&\hbar\left[  \frac{(R_{\mathrm{K}}/R)\nu_{F}SL}{2\pi(k_{B}T)^{2}}\right]  ^{1/3},\nonumber\\
\tau_{H}  &=& \frac{3\nu e^{2}RS}{L}\left(  \frac{\phi_{0}}{2\pi wB}\right) ^{2},\nonumber
\end{eqnarray}
where $\phi_{0}=h/e\simeq4.1\times10^{-15}$~T$\cdot$m$^{2}$ is the flux
quantum, $\mathrm{Ai}(x)$ is the Airy function and $\mathrm{Ai^{\prime}}(x)$
its derivative. The time $\tau_{N}$ is often called Nyquist time in reference
to the fluctuation-dissipation theorem used to evaluate the electromagnetic
fluctuations for the calculation of weak localization corrections.

Since expression~(\ref{WLairy}) includes electron-electron interactions, it
should be possible to deduce the contribution $\tau_{\mathrm{ee}}$ of the
screened Coulomb interaction on the phase coherence time. This can be done by
pointing out that
\begin{equation}
\frac{\mathrm{Ai}(x)}{\mathrm{Ai^{\prime}}(x)}=\frac{-1}{\sqrt{1/2+x}%
}(1+\epsilon(x)),
\end{equation}
where $|\epsilon(x)|<0.04$ for $x>0$. In practice, the experimental resolution
is not sufficient to distinguish a relative discrepancy smaller than 4\% of
the amplitude of weak localization corrections, which are themselves smaller
than 1\% of the measured signal. Hence we can write
\begin{equation}
\frac{\Delta R}{R}(B,T)=\frac{2R}{R_{K}L}\sqrt{\frac{D}{1/2\tau_{N}+1/\tau_{H}}}.
\end{equation}
A comparison with Eq.~(\ref{WeakLoc2}) (neglecting spin-orbit coupling) allows
us to extract the phase coherence time when it is limited by electron-electron
interactions:
\begin{eqnarray}
\tau_{\mathrm{ee}}  &=& \hbar\left[  \frac{(4/\pi)(R_{\mathrm{K}}/R)\nu_{F}%
SL}{(k_{B}T)^{2}}\right]  ^{1/3},\label{tauphiEEappendix}\\
&=& 2\tau_{N}.\nonumber
\end{eqnarray}
This expression of the phase coherence time $\tau_{\mathrm{ee}}$ is larger by
a factor $4/\pi^{2/3}\simeq1.9$ than the cut-off-dependent estimation in
Eq.~(\ref{tauphiEE_Keps}).

\section{Appendix C: Magnetic field dependence of spin-flip scattering}

This appendix present a simple calculation of electron spin-flip scattering
from magnetic impurities as a function of applied magnetic field $B$. The
calculation is carried out at first order in spin-flip scattering, neglecting
the Kondo effect. Moreover we consider here, for simplicity, magnetic
impurities of spin $1/2$.

The spin flip rate $\tau_{\mathrm{sf}}^{-1}(E,B)$ of an electron at energy $E$
is obtained from the Fermi Golden Rule
\begin{eqnarray}
\tau_{\mathrm{sf}}^{-1}(E,B)  &=& c_{\mathrm{mag}}\lambda\left\{  P_{-}(1-f_{T}(E-g\mu B))\right.  \nonumber\\
&& \left.  +P_{+}(1-f_{T}(E+g\mu B))\right\}  ,\label{tausfFGR}
\end{eqnarray}
where $c_{\mathrm{mag}}$ is the concentration of magnetic impurities, $\lambda$ is proportional to the modulus square
of the interaction potential electron-magnetic impurity and $P_{\pm}$ is the probability to have the magnetic impurity
in the up ($+$) or down ($-$) state relative to the magnetic field direction $B$. In absence of Kondo effect $\lambda$
is approximated as independent of energy and magnetic field.

Since at thermal equilibrium $P_{\pm}= f_{T}(\pm g \mu B)$, we obtain
\begin{equation}
\label{tausfEBT}\tau_{\mathrm{sf}}^{-1}(E,B) =
\frac{c_{\mathrm{mag}}\lambda(1+\exp(E/k_{B}T))/2}{\cosh(E/k_{B}T)+\cosh(g\mu B/k_{B}T)}.
\end{equation}

The spin-flip rate $\tau_{\mathrm{sf}}^{-1}(B)$ is averaged over electronic
excitations
\begin{equation}
\tau_{\mathrm{sf}}^{-1}(B)=\int_{-\infty}^{+\infty}dE\frac{f_{T}(E)(1-f_{T}(E))}{k_{B}T}\tau_{\mathrm{sf}}^{-1}(E,B),\nonumber
\end{equation}
which gives
\begin{equation}
\label{spinflipcrossover}\frac{\tau_{\mathrm{sf}}(B=0)}{\tau_{\mathrm{sf}}(B)}
= \frac{g \mu B / k_{B} T}{\sinh(g \mu B / k_{B} T)}.
\end{equation}

This result, also given in \cite{Meyer}, is a finite-temperature
generalization of the expression used by Benoit \textit{et al.}\cite{Benoit} A
rigorous theoretical calculation of the Aharonov-Bohm oscillation amplitude
$\Delta G_{h/e}$ in presence of magnetic impurities under a large externally
applied magnetic field was first presented by Fal'ko.\cite{FalkoAB}. A
complete derivation of the magnetic field dependence of $\Delta G_{h/e}$ from
first principles was finally published recently by Vavilov and
Glazman.\cite{VavilovGlazman} As discussed in Section~VII, the Vavilov-Glazman
crossover function for $S=1/2$ is nearly indistinguishable from ours.


\begin{thebibliography}{9}
\bibitem {meso}For a review, see \textit{Mesoscopic Phenomena in Solids},
edited by B.L. Altshuler, P.A. Lee, and R.A. Webb, (Elsevier, Amsterdam, 1991).

\bibitem {phonons}A. Schmid, Z. Physik \textbf{259}, 421 (1973).

\bibitem {AAK}B.L. Altshuler, A.G. Aronov, and D.E. Khmelnitsky, J. Phys. C
\textbf{15}, 7367 (1982).

\bibitem {AA}B.L. Altshuler and A.G. Aronov, in \textit{Electron-Electron
Interactions in Disordered Systems,} edited by A.L. Efros and M. Pollak (Elsevier Science Publishers B.V., 1985).

\bibitem {Wind}S. Wind, M.J. Rooks, V. Chandrasekhar, D.E. Prober, Phys. Rev.
Lett. \textbf{57}, 633 (1986).

\bibitem {Echternach}P.M. Echternach, M.E. Gershenson, H.M. Bozler, A.L.
Bogdanov, and B. Nilsson, Phys. Rev. B \textbf{48}, 11516 (1993).

\bibitem {MJW}P. Mohanty, E.M.Q. Jariwala, and R.A. Webb, Phys. Rev. Lett.
\textbf{78}, 3366 (1997).

\bibitem {PRLrelax}H. Pothier, S. Gu\'{e}ron, N.O. Birge, D. Esteve, and M.H.
Devoret, Phys. Rev. Lett. \textbf{79}, 3490 (1997).

\bibitem {names}Some of the measurements shown in this article have already
been published elsewhere.\cite{Gougam,PierrePesc,PierrePHD,PierreAB} (A list
of samples with cross-references to previous articles is given in
Ref.~\cite{correspondance}.)

\bibitem {adhesion}We did not need to perform ion-etching of the
surface\cite{MJW} or pre-deposition of another metallic layer to promote the
adhesion of the films.

\bibitem {PierrePHD}F.~Pierre, Ann. Phys. (Paris) \textbf{26}, N4 (2001).

\bibitem {NoteEVAP}For most samples we first melt the source material,
evaporate 10-20~nm with a shutter protecting the sample, and pump down the
chamber again for about 15~mn before the actual deposition. This
pre-evaporation covers the walls of the evaporator with a clean layer of metal
and makes it possible to maintain the pressure below $10^{-6}$~mbar during
material deposition. We did not notice a reproducible difference between
samples for which we did or did not follow this procedure.

\bibitem{WLreview} G. Bergmann, Phys. Rep. \textbf{107}, 1 (1984); S.
Chakravarty and A. Schmid, Phys. Rep. \textbf{140}, 19 (1986).

\bibitem{correspondance} Ag(6N)a is Ag in Ref. \cite{Gougam}, and Ag1 in Ref.
\cite{PierrePHD}. Ag6N(b) is Ag2 in Ref. \cite{PierrePHD}. Au(6N) is AuMSU in Ref. \cite{PierrePesc,PierrePHD}. Cu(6N)a
to Cu(6N)d are Cu1 to Cu4 in Ref. \cite{PierreAB}. Cu(5N)a is Cu in Ref. \cite{Gougam}, and Cu1 in Ref.
\cite{PierrePHD}. Cu(5N)b is Cu2 in Ref. \cite{PierrePHD}.

\bibitem{AleinerWav} I.L. Aleiner, B.L. Altshuler, and M.E. Gershenson, Waves
Random Media \textbf{9}, 201 (1999).

\bibitem {AAwl}B.L. Altshuler et A.G. Aronov, Pis'ma Zh. Eksp. Teor. Fiz.
\textbf{33}, 515 (1981) [JETP Lett. \textbf{33}, 499 (1981)].

\bibitem{NoteWidth} Several explanations can account for the small difference,
mostly observed on silver samples, between the width measured using an electronic microscope and the one used to make
accurate fits of low field magnetoresistance with weak localization theory. Amongst them we point out the
non-uniformity of our wires, our limited accuracy on the average width (about $\pm5$~nm), the fact we assume a
rectangular section for the wires, and most likely the limit of the diffusive approximation in silver wires for which
the elastic mean free path is not always very small compared to $w$ and which exhibit visible grain structure.

\bibitem {NoteAu}Samples made of our 4N gold source also display anomalous
dephasing. However, we could detect in these samples the presence of about
50~ppm of magnetic impurities, probably iron, that explains the observed
temperature dependence of $\tau_{\phi}$. For details see.\cite{PierrePesc}

\bibitem {AAG}B.L. Altshuler, M.E. Gershenson, and I.L. Aleiner, Physica E
\textbf{3}, 58 (1998).

\bibitem {dephasingMI}S. Hikami, A.I. Larkin, and Y. Nagaoka, Prog. Theor.
Phys. \textbf{63}, 707 (1980).

\bibitem {IFS}Y. Imry, H. Fukuyama, and P. Schwab, Europhys. Lett.
\textbf{47}, 608 (1999).

\bibitem {2CK}A. Zawadowski, J. von Delft, and D.C. Ralph, Phys. Rev. Lett.
\textbf{83}, 2632 (1999).

\bibitem {GZ}D.S. Golubev and A.D. Zaikin, Phys. Rev. Lett. \textbf{81}, 1074 (1998).

\bibitem {Gougam}A.B. Gougam, F. Pierre, H. Pothier, D. Esteve, and N.O.
Birge, J. Low Temp. Phys. \textbf{118}, 447 (2000).

\bibitem {PierreAB}F. Pierre and N.O. Birge, Phys. Rev. Lett. \textbf{89},
206804 (2002); F.~Pierre and N.O.~Birge, to appear in J. Phys. Soc. Jpn.

\bibitem {Haesendonck}J. Vranken, C. Van Haesendonck, and Y. Bruynseraede,
Phys. Rev. B \textbf{37}, 8502 (1988).

\bibitem {Wohlleben}D.K. Wohlleben and B.R. Coles, \textit{Magnetism}, edited
by H. Suhl (Academic, New York, 1973), Vol. 5.

\bibitem {noteIonImp}The energy of the ions, 70~keV, was chosen to obtain an
homogeneous concentration of manganese atoms in the thickness of the wires.

\bibitem {Falko}V.I. Fal'ko, JETP Lett. \textbf{53}, 340 (1991).

\bibitem {Maple}M.B. Maple, \textit{Magnetism}, edited by H. Suhl (Academic,
New York, 1973), Vol. 5.

\bibitem {Haesendonck_NS}C. Van Haesendonck, J. Vranken, and Y. Bruynseraede,
Phys. Rev. Lett. \textbf{58}, 1968 (1987).

\bibitem {noteSpinFit}In vacuum, the magnetic moment of manganese atoms is $S=5/2$, but $S=5/2$ does not allow accurate
fits of the measurements. It has been suggested that the random crystal field effectively reduces the degeneracy of the
lowest-lying spin multiplet to a doublet, regardless of the actual spin of the magnetic impurity (L. Glazman, private
communication). See also O. Ujsaghy, A. Zawadowski, and B.L. Gyorffy, Phys. Rev. Lett. \textbf{76}, 2378 (1996)

\bibitem {rhoKondo}A.C. Hewson, \textit{The Kondo Problem to Heavy Fermions}
(Cambridge University Press, 1993).

\bibitem {rhoKondoB}The value of $B_{K}$ for low concentrations of iron in
gold can be found for instance in P. Mohanty and R.A. Webb, Phys. Rev. Lett.
\textbf{84}, 4481 (2000). This parameter is expected to be similar for other
Kondo impurities in silver and copper.

\bibitem {GZS}D.S. Golubev, A.D. Zaikin, and G. Sch\"{o}n, J. Low Temp. Phys.
\textbf{126}, 1355 (2002).

\bibitem {controversyGZS}see J. von Delft, cond-mat/0210644 and references therein.

\bibitem {Natelson}D. Natelson, R.L. Willett, K.W. West, and L.N. Pfeiffer,
Phys. Rev. Lett. \textbf{86}, 1821 (2001). In their article, the quantity Natelson \emph{et al.} refer to as
$\tau_\phi$ or $\tau_N$ is four times smaller than the phase coherence time extracted from Eq.~(\ref{WeakLoc2}) (see
Eq.~(1) in Natelson \emph{et al.}). We took this correction into account in Fig.~\ref{FigZaikin}. Note that the
diffusion coefficients used by Natelson \emph{et al.} were obtained from the conductivity of co-evaporated thin films.
Using the expected geometry and the measured resistance of the wires, as was done for all other samples, would give
diffusion coefficients about twice smaller, which would move the open circles in Fig.~\ref{FigZaikin} about one decade
to the right, and slightly upward.

\bibitem {GZ_PRB}D.S. Golubev and A.D. Zaikin, Phys. Rev. B \textbf{59}, 9195 (1999).

\bibitem {Anthore}A. Anthore, F. Pierre, H. Pothier and D. Esteve, Phys. Rev. Lett. \textbf{90}, 076806 (2003); A. Anthore, F. Pierre, H. Pothier, D. Esteve, and M.H. Devoret, in \emph{Electronic
Correlations: From Meso- to Nano-Physics}, Ed. T. Martin, G. Montambaux and J. Tr\^{a}n Thanh V\^{a}n (EDP Sciences,
2001); cond-mat/0109297.

\bibitem {AleinerTLS}I.L. Aleiner, B.L. Altshuler, and Y.M. Galperin, Phys.
Rev. B, \textbf{63}, 201401, (2001).

\bibitem {TLS_TK_Aleiner}I.L. Aleiner, B.L. Altshuler, Y.M. Galperin, and T.A.
Shutenko, Phys. Rev. Lett. \textbf{86}, 2629 (2001); I.L. Aleiner and D.
Controzzi, Phys. Rev. B \textbf{66}, 045107 (2002).

\bibitem {TLS_TK_Zawa}A. Zawadowski and G. Zarand, cond-mat/0009283; O.
Ujsaghy, K. Vladar, G. Zarand, and A. Zawadowski, J. Low Temp. Phys.
\textbf{126}, 1221 (2002).

\bibitem {Meyer}J.S. Meyer, V.I. Fal'ko, and B.L. Altshuler,
\textit{Strongly-Correlated Fermions and Bosons in Low-Dimensional Disordered
Systems}, edited by I. Lerner, B.L. Altshuler, V.I. Fal'ko and T. Giamarchi
(Kluwer Academic, 2002); cond-mat/0206024.

\bibitem {NoteAB}A.G. Aronov and Y.V. Sharvin, Rev. Mod. Phys. \textbf{59},
755 (1987); V. Chandrasekhar, Ph.D. thesis, Yale University (1989). We neglect here the triplet contribution to AB
oscillations since the spin-orbit length is much smaller than $L_{\phi}$.

\bibitem {NoteABnoise} At 40~mK and low magnetic field, our experimental noise floor dominates the amplitude of conductance fluctuations in the $h/e$ frequency window.
This explains the smaller relative increase of $\Delta G_{AB}$ at 40~mK compared to 100~mK.

\bibitem {ABfitnote}In principle, the constant $C$ in Eq.~(\ref{GAB}) is
calculable from the known sample geometry. In practice, the calculation is quite difficult in 4-terminal measurement
geometry (Thomas Ludwig, private communication). See D.P. DiVincenzo and C.L. Kane, Phys. Rev. B \textbf{38}, 3006
(1988), C.L. Kane, P.A. Lee and D.P. DiVincenzo, Phys. Rev. B \textbf{38}, 2995 (1988).

\bibitem {VavilovGlazman}M.G. Vavilov and L.I. Glazman, Phys. Rev. B \textbf{67}, 115310 (2003).

\bibitem {VG_S}Although the magnetic field dependence of the Vavilov-Glazman
prediction depends on the spin $S$ of the magnetic impurity, the quality of
the fit adjusting the parameter $g$ hardly changes with $S$, indicating that
the noise in the data is too large to permit a reliable determination of $S$
in our samples.

\bibitem {Shirane}B.X. Yang, J.M. Tranquada, and G. Shirane, Phys. Rev. B
\textbf{38}, 174 (1988).

\bibitem {PierrePesc}F. Pierre, H. Pothier, D. Esteve, M.H. Devoret, A.
Gougam, and N.O. Birge, in \textit{Kondo Effect and Dephasing in Low-Dimensional Metallic Systems}, edited by V.
Chandrasekhar, C. Van Haesendonck, and A. Zawadowski, (Kluwer, Dordrecht, 2001), p.~119; cond-mat/0012038.

\bibitem {PierreJLTP}F. Pierre, H. Pothier, D. Esteve, and M.H. Devoret, J.
Low Temp. Phys. \textbf{118}, 437 (2000).

\bibitem{KaminskyGlazman} A. Kaminski and L.I. Glazman, Phys. Rev. Lett.
\textbf{86}, 2400, (2001).

\bibitem {JJLin}It is not clear that the saturation of $\tau_{\phi}$\ observed in strongly disordered alloys, \emph{cf.} J.J. Lin, Y.L. Zhong, and T.J. Li, Europhys. Lett. \textbf{57}, 872
(2002), can be explained along the same lines.

\bibitem{HennySchonenberger} M. Henny, H. Birk, R. Huber, C. Strunk, A. Bachtold, M. Kr\"{u}ger, and C. Sch\"{o}nenberger., Appl. Phys. Lett. \textbf{71}, 773 (1997).

\bibitem{Wellstood} F.C. Wellstood, C. Urbina, and J. Clarke, Phys. Rev. B
\textbf{49}, 5942 (1992).

\bibitem {sigma}Three parameters describing effects of electron-phonon
scattering are used in this review: $B$ in Eq.~(\ref{fitee-eph}), $\Sigma$ and
$\kappa_{\mathrm{ph}}$ in Appendix~A. They are related by the
relations\cite{PierrePHD} $\Sigma=24\zeta(5)\nu_{F}\kappa_{\mathrm{ph}}%
k_{B}^{5}$ and $B=6\zeta(3)\kappa_{\mathrm{ph}}k_{B}^{3}$, with $\zeta(x)$ the Riemann zeta function:
$\zeta(5)\simeq1.04$ and $\zeta(3)\simeq1.2.$ Introducing the heat capacity coefficient $\gamma,$ one has the relation
$\Sigma\simeq1.05\gamma B.$

\bibitem {AndreevKamenev}A. Kamenev and A. Andreev, Phys. Rev. B \textbf{60},
2218 (1999).

\bibitem {noteReplaceCutoff}We replaced the cut-off at $\hbar/\tau_{\phi}(T)$
by $\hbar/\tau_{\mathrm{ee}}(T)$ whereas, when another inelastic process such as electron-phonon scattering limits the
quantum coherence, the integral should be cut-off at $\hbar/\tau_{\phi}$ ($>\hbar/\tau_{\mathrm{ee}}$). We neglect this
correction here since it applies only when the contribution of electron-electron interactions is already weak.

\bibitem {Benoit}A.D. Benoit, S. Washburn, R.A. Webb, D. Mailly, and L.
Dumoulin, \textit{Anderson Localization}, edited by T. Ando and H. Fukuyama
(Springer, 1988).

\bibitem {FalkoAB}V.I. Fal'ko, J. Phys: Condens. Matter \textbf{4}, 3943 (1992).

\end{thebibliography}
\end{document}